\documentclass[12pt,a4paper]{article}
\usepackage{amsmath}
\usepackage{graphicx}
\usepackage{enumerate}
\usepackage[authoryear,round]{natbib}
\setcitestyle{maxnames=3}
\usepackage{url} 

\usepackage{color}
\usepackage{amssymb}
\usepackage{amsfonts}
\usepackage{amsthm}
\usepackage{multirow}
\usepackage{float}
\usepackage{hyperref}
\usepackage{booktabs}
\usepackage{bm}
\usepackage{bbm}
\usepackage{setspace}
\usepackage{footmisc}
\usepackage{tikz}

\newcommand{\circled}[1]{%
	\tikz[baseline=(char.base)]\node[draw, shape=circle, inner sep=1pt](char){#1};%
}

\numberwithin{equation}{section}

\newtheorem{theorem}{Theorem}
\newtheorem{lemma}[theorem]{Lemma}
\newtheorem{corollary}[theorem]{Corollary}
\newtheorem{remark}[theorem]{Remark}
\newtheorem{definition}[theorem]{Definition}

\theoremstyle{definition}
\newtheorem{assumption}{Assumption}

\def\E{\mathbb{E}}
\def\P{\mathbb{P}}

\usepackage[ruled,vlined]{algorithm2e}

\SetAlgoNoEnd
\DontPrintSemicolon
\SetKwInOut{Output}{Output}

\begin{document}

\def\spacingset#1{\renewcommand{\baselinestretch}%
{#1}\small\normalsize} \spacingset{1.2}


\title{\bf Robust Semiparametric Inference for Bayesian Additive Regression Trees}
\author{Christoph Breunig\thanks{University of Bonn. Email: \url{cbreunig@uni-bonn.de}}\and Ruixuan Liu\thanks{Chinese University of Hong Kong. Email: \url{ruixuanliu@cuhk.edu.hk}}\and Zhengfei Yu\thanks{University of Osaka. Email: \url{yu.zhengfei@osaka-u.ac.jp}}}
\maketitle

\medskip
\begin{abstract}
\spacingset{1.2}
Bayesian additive regression trees (BART) provide a class of flexible nonparametric regressions in machine learning. Their rich tree-based structure can generate non-Donsker classes of regression functions. We study posterior inference for the population mean under missing at random. We derive a new Bernstein--von Mises (BvM) theorem showing that even a one-step posterior based on BART and Bayesian-bootstrap reweighting retains a non-negligible bias term in the non-Donsker regime. To remove this term, we introduce RoBART, a posterior correction that adjusts each draw using pilot estimators of the outcome regression and propensity score. We prove a BvM theorem for the corrected posterior and develop a cross-fitted version using fold-specific posteriors. We provide primitive assumptions tailored to BART. We show that the posterior mean coincides exactly with the conventional augmented inverse-probability-weighted (AIPW) estimator and, under cross-fitting, with the corresponding double machine learning (DML) estimator. RoBART therefore provides a posterior distribution with asymptotically valid uncertainty quantification around this well-established frequentist center. Simulations and an empirical illustration demonstrate competitive finite-sample performance relative to existing methods.
\end{abstract}

\noindent%
\vfill

\newpage
\spacingset{1.2}

\section{Introduction}\label{sec:intro}

 Bayesian additive regression trees (BART) is one of the most powerful statistical methods \citep{chipman2010bart}. Its empirical success in predictive modeling has been demonstrated in numerous studies. In the context of nonparametric function estimation, BART achieves adaptive, near rate-minimax posterior contraction, as shown by \cite{linero2018bart,rockova2020posterior,rockova2020bart,jeong2023art}. Beyond prediction, BART has also shown strong performance in semiparametric inference for causal parameters of interest \citep{dorie2019auto,hahn2020bart}. However, a theoretical justification for semiparametric inference based on BART under reasonable regularity conditions is still developing.

The difficulty arises partly from the structure of tree-based methods. BART represents regression functions through ensembles of piecewise-constant trees and is therefore intrinsically nonsmooth. For Hölder-type function classes, standard Donsker conditions typically require smoothness exceeding one-half of the covariate dimension. Such conditions become increasingly restrictive in moderately high-dimensional settings and are particularly difficult to reconcile with piecewise-constant tree priors. At the same time, flexible nuisance estimation is essential in missing-data and causal-inference problems \citep{chernozhukov2017double,chernozhukov2018double,athey2021policy}, where misspecification of the outcome regression may lead to invalid inference. 
Our objective is to establish valid posterior inference for the finite-dimensional parameter of interest when the nuisance regression is modeled by BART in a non-Donsker regime where existing theory does not apply. 
While augmented inverse-probability-weighted (AIPW) estimation, combined with cross-fitting and flexible nuisance estimation as in double machine learning (DML), already provides robust frequentist point estimation in this setting, our focus is on valid posterior uncertainty quantification around the same center.

We consider estimation of the population mean under missing at random and subsequently extend the construction to average treatment effects (ATE). 
Starting from the AIPW representation, we first consider a baseline one-step posterior that combines a BART posterior for the conditional mean, a pilot estimator of the propensity score, and Bayesian-bootstrap weights. This construction is a specialization of the one-step posterior of \citet{yiu2023corrected}, obtained by replacing their posterior for the propensity score with a point mass at a pilot estimate. We use this one-step posterior as a benchmark for studying posterior inference with BART. 
We derive a semiparametric Bernstein–von Mises theorem for the resulting one-step posterior. 
The Bernstein--von Mises limit of the one-step BART posterior is generally shifted by an additional first-order bias term that need not vanish outside the Donsker regime. The same bias term arises in \cite{BLY2022} from a different posterior construction based on prior adjustment. Thus, two distinct posterior constructions encounter the same first-order obstruction. We characterize this obstruction for the one-step BART posterior and remove it by a draw-specific correction.

To remove this term, we introduce robust BART, or RoBART, which applies an explicit correction to each posterior draw using pilot estimators of the conditional mean and propensity score. The correction is analytically simple and preserves the posterior variation generated by the BART posterior and the Bayesian bootstrap.  We establish a Bernstein--von Mises theorem for the corrected posterior, yielding $\sqrt{n}$-rate inference and the semiparametric efficiency bound under product-rate conditions that permit weaker regularity of the conditional mean function to be offset by stronger regularity of the propensity score, and vice versa.

An important justification for the correction comes from its posterior mean. Conditional on the observed data, the posterior mean of the corrected procedure coincides exactly with the AIPW estimator based on the pilot nuisance estimates. We also develop a cross-fitted version in which BART posteriors are constructed separately within each evaluation fold and the corresponding pilot estimators are trained on the complementary observations. The average of the fold-specific posterior means then coincides exactly with the cross-fitted AIPW estimator \citep{robins1994regression}, which, when combined with machine-learning nuisance estimators and cross-fitting,  is the DML estimator  \citep{chernozhukov2018double}. Thus, RoBART does not introduce a different point estimator from DML. Rather, it provides a corrected posterior distribution for uncertainty quantification around the same semiparametrically robust center.

Our theoretical results cover both the construction of pilot estimators using an auxiliary sample and a cross-fitted procedure that uses the full sample. In contrast to standard empirical-process arguments, the analysis does not require the nuisance-function class to be Donsker. This feature is particularly relevant for BART because its piecewise-constant structure limits the smoothness that can be exploited directly. The conditions instead allow the regularity of the outcome regression and propensity score to compensate for one another through the familiar product-type rate restriction.

Our RoBART procedure is best viewed as a calibrated Bayesian construction in the sense of \cite{rubin1984bayesian}. It begins with a prior and a posterior for the nuisance function, but augments the resulting posterior draws with a correction designed to deliver valid frequentist inference. This perspective is also related to the notion of pragmatic Bayesian procedures discussed by \citet{ritov2014}. The resulting credible intervals retain a posterior interpretation conditional on the observed data, while the Bernstein–von Mises theorem establishes their asymptotic frequentist validity.

Admittedly, the use of pilot estimation and cross-fitting deviates from the Bayesian principle in the strict sense. The construction is closely related to the active DML literature \citep{chernozhukov2017double,chernozhukov2018double}. DML combines debiased moment conditions with cross-fitting to permit flexible nuisance estimation without relying on restrictive Donsker conditions. RoBART shares this robustness principle but uses the posterior distribution generated by a nonparametric Bayesian learner to conduct inference. The exact equivalence of the posterior mean and the corresponding DML estimator clarifies the relationship between the two approaches: they share the same point-estimation center, whereas RoBART uses the corrected BART posterior and Bayesian-bootstrap variation to quantify uncertainty around that center.

Monte Carlo experiments illustrate the more robust performance of RoBART relative to other Bayesian and frequentist methods.
In increasingly complex designs, standard BART and the one-step BART 
 can exhibit substantial undercoverage, whereas RoBART maintains coverage closer to the nominal level. Because the posterior mean of RoBART coincides with the corresponding DML estimator,  the comparison between these two focuses on uncertainty quantification.  Using the same folds and nuisance estimates,  RoBART yields better coverage performance than DML,  especially for more complex designs.  We also apply RoBART in an empirical illustration.

 \textit{Related literature.} 
\cite{ray2020causal} use adjusted priors to weaken regularity requirements
in semiparametric Bayesian inference. \cite{BLY2022} combine a related data-dependent prior adjustment with
an explicit posterior correction and obtain doubly robust Bayesian inference for the ATE. The present paper addresses a different problem: inference when the outcome regression is given by a BART posterior
whose function class need not satisfy the Donsker conditions used in existing semiparametric Bayesian asymptotic theory. In addition, the prior adjustment is difficult to incorporate into BART-based modeling of the conditional outcome. Starting
from the one-step posterior of \cite{yiu2023corrected}, we characterize the non-negligible bias that appears in
this regime. We show that it coincides with the bias generated by the distinct prior adjustment construction of \cite{BLY2022}, and develop a draw-wise correction device to remove it. For the cross-fitted procedure, we establish a strong equivalence result that shows the posterior mean after our correction coincides exactly with the corresponding AIPW/DML estimator, which is new to the literature. The technical contribution is a BvM theorem that justifies the uncertainty quantification from the corrected posterior distribution under primitive conditions tailored to BART. 
 
%
 
 As a general nonparametric regression method, BART achieves adaptive, near rate-minimax posterior contraction, a major breakthrough due to \cite{linero2018bart,rockova2020posterior,rockova2020bart,jeong2023art}. Compared with the extensive results on the posterior convergence for BART, the corresponding inferential theory is relatively scarce. \cite{rockova2020bart} establish a semiparametric BvM theorem using BART-type priors for a particular type of linear functional of the regression function under the fixed design. That result requires the Donsker property of the conditional mean. For Bayesian CART (a single tree rather than a forest), \cite{castillo2021CART} also establish a general semiparametric BvM theorem within the Donsker regime.
 
  \textit{Organization.} The paper is structured as follows. Section \ref{sec:model} describes the missing-data model together with the identifying assumptions and presents the robust semiparametric Bayesian procedure. We highlight the strong equivalence between our point estimator and the AIPW (or DML) estimator. In Section \ref{sec:asympt}, we derive the BvM theorem under high-level assumptions using pilot estimators from auxiliary data. Section \ref{sec:bart} establishes corresponding results for the cross-fitted version under more primitive conditions for BART. Section \ref{sec:numerical} provides numerical results via Monte Carlo simulations and an empirical illustration. 
 Proofs of the main theoretical results are presented in Appendix \ref{appendix:main:proofs}, and the lemmas and auxiliary results used therein are collected in Appendix \ref{appendix:auxiliary}. Appendix \ref{appendix:other_para} extends the framework to inference on average treatment effects. Appendix \ref{sec:barttheory} contains the posterior contraction and empirical process results for regression trees that are used in the proof of Theorem \ref{thm:BvM:BART}, and Appendix \ref{appendix:implem} provides implementation details of BART.

 \textit{Notation.}  We adopt the standard empirical process notation as follows. For a function $h$ of a random vector $Z$ that follows distribution $P$, we let $P[h]=\int h(z)\,\mathrm{d} P(z),\mathbb{P}_n[h]=n^{-1}\sum_{i=1}^{n}h(Z_i)$, and $\mathbb{G}_n[h]=\sqrt n\left(\mathbb{P}_n-P\right)[h]$. We further write $\mathbb{P}^*_n[h]=\sum_{i=1}^{n}W_{ni}h(Z_i)$ for the bootstrap empirical measure induced by the Bayesian-bootstrap weights $W^{(n)}=(W_{n1},\ldots,W_{nn})^\top$.
 For two sequences $\{a_n\}$ and $\{b_n\}$ of positive numbers, we write $a_n \lesssim b_n$ if $\limsup_{n\to\infty} (a_n / b_n)<\infty$, and $a_n \sim b_n$ if $a_n \lesssim b_n$ and $b_n \lesssim a_n$.
	\section{Setup and Implementation}\label{sec:model}
This section provides the main setup of our missing-data framework and motivates the new Bayesian methodology. We provide the implementation of our robust Bayesian algorithm and discuss its connections to and differences from the existing literature.  

\subsection{The Model Setup}
We consider a family of probability distributions $\{P_\eta:\eta\in\mathcal H\}$ for some parameter space $\mathcal H$. The (possibly infinite-dimensional) parameter $\eta$ characterizes the probability model. An independent and identically distributed (i.i.d.) sample $\{(R_iY_i,R_i,X_i^\top)^\top\}_{i=1}^n$ is available, where the dependent variable $Y_i$ is observed only if $R_i=1$ and it is missing if $R_i=0$. 
We assume the outcomes $Y_i$ are missing at random (MAR); that is,  the outcome $Y_i$ and the missingness indicator $R_i$ are conditionally independent given $X_i$. 

Throughout the paper, the distribution of the observable vector $(RY,R,X^\top)^\top$ is modeled by three components: (i) the marginal distribution $\mu_{\eta}(x)$ of the $p$-dimensional covariates $X$; (ii) the propensity score $\pi_{\eta}(x)=P_{\eta}(R=1\mid X=x)$; and (iii) the conditional density function $f_{\eta}(\cdot \mid x)$ of the outcome $Y$ given $X=x$ and $R=1$. Consequently, the conditional mean function of the outcome given covariates is
\begin{align*}
	m_\eta(x)=\int yf_\eta(y\mid x)\,\mathrm{d}y.
\end{align*} 
Under the MAR assumption, the probability density function $p_{\eta}$ for $Z_i=(R_iY_i,R_i,X_i^\top)^\top$ can be written as
\begin{equation}\label{def:density}
	p_{\eta}(z)=\frac{\mathrm{d} \mu_{\eta}(x)}{\mathrm{d} x}\pi^r_{\eta}(x)(1-\pi_{\eta}(x))^{1-r}f^r_{\eta}(y\mid x).
\end{equation}
Let $\eta_0$ be the true value of the parameter and denote $P_0=P_{\eta_0}$, which corresponds to the frequentist distribution that generates the observed data. Accordingly, we denote $\mu_0$ as the true marginal covariates distribution, $\pi_0$ as the true propensity score, and $f_{0}(\cdot\mid x)$ as the true conditional density of $Y$ given $X=x$ and $R=1$.

The key parameter of interest is the mean of the outcome variable:
\begin{equation*}
\chi_0:=\mathbb{E}_0[Y_i],
\end{equation*}
where $\mathbb{E}_0[\cdot]$ denotes expectation under $P_0$. Under the MAR assumption, $\chi_0$ can be identified using the conditional mean function $m_0(x):=\mathbb{E}_0[Y_i\mid R_i=1,X_i=x]$. Specifically, we have
\begin{equation*}
\chi_0=\int m_0(x)\,\mathrm{d}\mu_0(x),
\end{equation*}
where $\mu_0$ denotes the marginal distribution of $X_i$ under $P_0$. For our approach, it is useful to consider the following augmented representation of $\chi_0$:
\begin{equation}\label{def:chi0}
\chi_0=\int\left(m_0(x)+\gamma_0(r,x)\bigl(y-m_0(x)\bigr)\right)\,\mathrm{d}P_0(y,r,x),
\end{equation}
where the Riesz representer is given by
\begin{equation}\label{riesz:def}
\gamma_0(r,x)=\frac{r}{\pi_0(x)},
\end{equation}
with $\pi_0(x):=\mathbb{P}_0(R_i=1\mid X_i=x)$. In particular, $\mathbb{E}_0[m_0(X_i)\gamma_0(R_i,X_i)]=\chi_0$; see \citet{chernozhukov2018double}.

Let $W^{(n)}:=(W_{n1},\ldots,W_{nn})$ denote the Bayesian-bootstrap weights of \citet{rubin1981bayesian}, defined by
$W_{ni}=e_i/\sum_{j=1}^n e_j$ for $e_i \stackrel{iid}{\sim} \textup{Exp}(1)$, $i=1,\dots,n$.
Motivated by the augmented representation in \eqref{def:chi0}, we define the Bayesian-bootstrap corrected draw
\begin{equation}\label{OneStep}
\chi_{\eta, \widehat \gamma}:=\sum_{i=1}^n W_{ni}\left(m_\eta(X_i)+\widehat{\gamma}(R_i,X_i)\bigl(Y_i-m_\eta(X_i)\bigr)\right),
\end{equation}
where $m_\eta(\cdot)$ is a stochastic function endowed with a prior and
\begin{equation*}
\widehat{\gamma}(r,x):=\frac{r}{\widehat{\pi}(x)}
\end{equation*}
is the estimated Riesz representer based on a pilot estimator $\widehat{\pi}$ of $\pi_0$ computed using an auxiliary data set. The conditional distribution of $\chi_{\eta,\widehat\gamma}$ given the observed data $Z^{(n)}:=(Z_1,\ldots,Z_n)$ is jointly induced by the posterior law of $m_\eta$, denoted by $\Pi_m(\cdot\mid Z^{(n)})$, and the distribution of the Bayesian-bootstrap weights $W^{(n)}$. The latter distribution is known and straightforward to simulate. Following the notation for bootstrap laws used by \citet{van1996empirical}, we denote it by $\Pi_W(\cdot\mid Z^{(n)})$. We refer to the resulting conditional law of $\chi_{\eta, \widehat \gamma}$ as the baseline one-step posterior for $\chi_0$. While our general results apply to a broad class of nonparametric priors for $m_\eta$, we particularly analyze BART priors under more primitive conditions.\footnote{Posterior draws of the vector $(m_\eta(X_1),\ldots,m_\eta(X_n))$ can be obtained using the R package $\mathsf{BART}$; see \citet{sparapani2021nonparametric}.}

To motivate this construction from a more conventional Bayesian perspective, suppose that $\eta$ indexes a probability model $P_\eta$ with corresponding conditional mean function $m_\eta$ and propensity score $\pi_\eta$. The associated model-based functional is
\begin{equation}\label{def:chi:eta}
\chi_\eta=\int\left(m_\eta(x)+\frac{r}{\pi_\eta(x)}\bigl(y-m_\eta(x)\bigr)\right)\,\mathrm{d}P_\eta(y,r,x).
\end{equation}
The conditional distribution of \eqref{OneStep} can be viewed as a modular approximation to the posterior distribution of \eqref{def:chi:eta}. In particular, replacing $\pi_\eta$ by the pilot estimator $\widehat{\pi}$ amounts to placing a degenerate distribution at $\widehat{\pi}$, while uncertainty about $m_\eta$ is retained through its nonparametric posterior. Related uses of preliminary propensity-score estimators in semiparametric Bayesian inference are considered by \citet{ray2020causal}. Alternatively, one could place an independent nondegenerate prior on the propensity score and propagate the resulting posterior uncertainty through the corrected posterior.

For the integrating probability measure, we use the Bayesian bootstrap. More precisely, the Bayesian bootstrap can be viewed as the noninformative limiting posterior arising from a sequence of Dirichlet process models as the prior concentration parameter tends to zero. 
Equation \eqref{OneStep} can be viewed as a specialization of the one-step posterior of \citet{yiu2023corrected}: uncertainty about $m_\eta$ is retained through its nonparametric posterior, the posterior for the propensity score is replaced by a point mass at a pilot estimator, and the integrating distribution is represented by the Bayesian bootstrap. We use this construction as the baseline one-step posterior. Section \ref{sec:mot} shows why, when the conditional mean is modeled by BART, this baseline generally requires an additional draw-wise correction outside the Donsker regime.

\subsection{Motivation of Robust Bayesian Procedure}\label{sec:mot}
Our construction is guided by the efficient influence function for the mean response under missing at random. For a generic nuisance parameter $\eta=(m_\eta,\pi_\eta)$, consider the influence-function map
\begin{align}\label{eif}
\widetilde\chi_{\eta}(Z)=m_{\eta}(X)+\gamma_\eta(R,X)\{Y-m_\eta(X)\}-\chi_{\eta},
\end{align}
where the corresponding Riesz representer is
\begin{align}\label{riesz:def2}
\gamma_\eta(R,X)=\frac{R}{\pi_\eta(X)}.
\end{align}
Evaluated at the true nuisance parameter $\eta_0=(m_0,\pi_0)$, \eqref{eif} is the efficient influence function for the target parameter. Let $\widehat\pi$ denote a pilot estimator of the propensity score based on the auxiliary sample, and define $\widehat\gamma(r,x)=r/\widehat\pi(x)$. The posterior correction also uses a pilot estimator $\widehat m$ of the conditional mean function $m_0$. In our implementation, $\widehat m$ is taken to be the posterior mean of a BART model fitted to the auxiliary sample.

Estimating the nuisance functions on independent auxiliary data simplifies the analysis by rendering certain empirical-process remainder terms asymptotically negligible. Sample splitting has a long history in semiparametric estimation \citep{schick1986semi,van1998asymptotic} and is central to the recent DML literature \citep{chernozhukov2018double}. Related auxiliary-sample constructions have also been used for propensity-score-adjusted Bayesian priors in missing-data problems \citep{ray2020causal}. The auxiliary-sample formulation allows us to present the main argument transparently, while the cross-fitted extension avoids the need for a separate sample.

We show that the one-step Bayesian procedure $\chi_{\eta, \widehat \gamma}$ defined in \eqref{OneStep}, which uses a pilot estimator of the propensity score, generally admits a Bernstein--von Mises approximation (see Theorem \ref{thm:BvM}) only after accounting for the additional centering bias
\begin{align*}
b_{0,\eta}&=\mathbb P_n\left[m_\eta(X)+\gamma_0(R,X) (Y-m_\eta(X))\right] -\mathbb P_n\left[m_0(X)+\gamma_0(R,X)(Y-m_0(X))\right] \\
&=\mathbb P_n\left[(1-\gamma_0(R,X))(m_\eta(X)-m_0(X))\right].
\end{align*}
This term need not be asymptotically negligible when the posterior for the outcome regression concentrates too slowly for standard Donsker arguments to apply.

The purpose of the posterior correction is to remove the draw-specific centering bias induced by nonparametric Bayesian modeling of the outcome regression, while preserving the posterior variation generated by the BART posterior and the Bayesian-bootstrap weights. The correction also has a useful point-estimation interpretation: it recenters the posterior at the conventional doubly robust estimator based on the pilot nuisance estimates.

Our  proposal considers an estimator of the bias term $b_{0,\eta}$ given by
\begin{align*}
\widehat b_\eta=\mathbb P_n\left[(1-\widehat\gamma(R,X))(m_\eta(X)-\widehat m(X))\right] ,
\end{align*}
where $\widehat m$ and $\widehat\gamma$ are pilot estimators computed from an auxiliary sample. The resulting corrected posterior draw is $\check\chi_\eta=\chi_{\eta, \widehat \gamma}-\widehat b_\eta$.  An important feature of this correction is that the conditional posterior mean of $\check\chi_\eta$ coincides exactly with the conventional AIPW estimator based on the pilot estimators.

To see this, let $\mathbb E_\Pi$ denote expectation with respect to the joint posterior distribution of $m_\eta$ and the Bayesian-bootstrap weights, conditional on $Z^{(n)}$, where the pilot estimators $\widehat m$ and $\widehat\gamma$ are treated as fixed. Since the Bayesian-bootstrap weights are independent of $m_\eta$ and satisfy $\mathbb E_\Pi[W_{ni}\mid Z^{(n)}]=1/n$, we obtain
\begin{align*}
\mathbb E_\Pi\left[ \chi_{\eta, \widehat \gamma}\mid Z^{(n)} \right] &= \mathbb P_n\left[ \mathbb E_\Pi[m_\eta(X)\mid Z^{(n)}] +
\widehat\gamma(R,X) \left( Y-\mathbb E_\Pi[m_\eta(X)\mid Z^{(n)}] \right) \right].
\end{align*}
Moreover, since $\widehat b_\eta$ is affine in $m_\eta$, we have
\begin{align*}
\mathbb E_\Pi\left[ \widehat b_\eta\mid Z^{(n)} \right] &=
\mathbb P_n\left[ \mathbb E_\Pi[m_\eta(X)\mid Z^{(n)}] + \widehat\gamma(R,X) \left( Y-\mathbb E_\Pi[m_\eta(X)\mid Z^{(n)}] \right) \right] \\
&\quad- \mathbb P_n\left[ \widehat m(X) + \widehat\gamma(R,X) (Y-\widehat m(X))\right].
\end{align*}
It follows that the posterior mean of the corrected draw
$\check\chi_\eta=\chi_{\eta, \widehat \gamma}-\widehat b_\eta$ is
\begin{align}\label{EquivalenceAIPW}
\mathbb E_\Pi\left[\check\chi_\eta\mid Z^{(n)}\right]
=\mathbb P_n\left[\widehat m(X)+\widehat\gamma(R,X)(Y-\widehat m(X))\right].
\end{align}
Thus, the conditional posterior mean of the corrected Bayesian procedure
coincides exactly with the conventional AIPW estimator based on
$(\widehat m,\widehat\pi)$. The BART posterior draws and Bayesian-bootstrap weights determine the posterior distribution around this center, but not the center itself. In the cross-fitted construction, averaging the fold-specific posterior means analogously yields the conventional cross-fitted AIPW, or DML, estimator.

\subsection{Robust Bayesian Procedure with Auxiliary Data}

We first present the implementation of our robust inference procedure under the assumption that an auxiliary sample is available for computing the pilot estimators. This auxiliary-sample formulation provides a convenient starting point for presenting the main idea. In the next subsection, we extend the approach to a $K$-fold cross-fitted version, which makes efficient use of the full data.

The inputs to the algorithm are the observable data 
$\{(R_i Y_i, R_i, X_i^\top)^\top\}_{i=1}^n$, 
the pilot estimators $\widehat{\gamma}$ and $\widehat{m}$ as discussed above, 
and the number of posterior draws $S$.
The posterior distribution of $(m_\eta(X_i))_{i=1}^n:=(m_\eta(X_1),\ldots, m_\eta(X_n))$ is generated by applying BART to estimate $m_{\eta}(\cdot)$ using the data $\{(Y_i,X_i^\top):R_i=1\}_{i=1}^n$.

\begin{algorithm}
    \caption{Posterior Computation}
    \label{algorithm}

    \For{$s=1,\ldots,S$}{

        (a) Obtain one draw from the posterior of $(m_\eta(X_i))_{i=1}^n$, and denote it by $(m_\eta^s(X_i))_{i=1}^n$. \;

        (b) Generate $W_{ni}^s = e_i^s / \sum_{j=1}^n e_j^s$ with $e_i^s \stackrel{iid}{\sim} \mathrm{Exp}(1)$ for all $1\leq i\leq n$.

        (c) Calculate $\check{\chi}_\eta^s= \chi_{\eta, \widehat \gamma}^s-\widehat{b}_\eta^s$, where
        \begin{equation} \label{recentered_bay_est_nocf}
            \chi_{\eta, \widehat \gamma}^s=\sum_{i=1}^n W_{ni}^s\left(m_\eta^s(X_i)+\widehat{\gamma}(R_i,X_i)
                \bigl(Y_i-m_\eta^s(X_i)\bigr)\right),
        \end{equation}
        and the debiasing term is
        \begin{equation}\label{debiased_bay_est_nocf}
            \widehat{b}_\eta^s=\frac{1}{n}\sum_{i=1}^n\left(\widehat{\gamma}(R_i,X_i)-1\right)
            \left( \widehat{m}(X_i)-m_\eta^s(X_i)\right).
        \end{equation}
        \vskip -.5cm
    }

    \Output{
        $\left\{\check{\chi}_\eta^s:s=1,\ldots,S\right\}$
    }
\end{algorithm}

Given the posterior simulation draws, the $100\cdot(1-\alpha)\%$ credible set $\mathcal{C}_n(\alpha)$ for the parameter of interest $\chi_0$ is computed by
\begin{equation}\label{CS:def}
	\mathcal{C}_n(\alpha)=\big\{\chi: q(\alpha/2)\leq \chi \leq q(1-\alpha/2)\big\},
\end{equation}
where $q(a)$ denotes the $a$-quantile of $\{\check{\chi}_\eta^{s}:s=1,\ldots,S\}$. We take the Bayesian point estimator (the posterior mean) by averaging the simulation draws: $\overline{\chi}_{\eta}=S^{-1}\sum_{s=1}^S \check{\chi}_\eta^{s}$.

\begin{remark}[Estimation of the Riesz Representer]
Our proposal builds on a plug-in approach that employs pilot estimators of the Riesz representer. 
Our approach is motivated by the following considerations. First, by adopting sample splitting for the pilot estimators, our proposal makes it possible to relax the Donsker condition.
Sample splitting has long been used in semiparametric models \citep{schick1986semi,van1998asymptotic} and has regained popularity in the DML literature \citep{chernozhukov2017double,chernozhukov2018double}. Second, in more complicated semiparametric models, the Riesz representer may lack a tractable analytical form, but one can still construct a pilot estimator following the approach of \cite{chernozhukov2017robust}. In such cases, designing a feasible algorithm to obtain its posterior can be challenging. 
Finally, compared to the conditional mean function, the Riesz representer is often of secondary interest; one can view the plug-in estimator as a particular type of degenerate prior for the propensity score \citep{ray2020causal}. 
\end{remark}

\subsection{Cross-fitted Robust Bayesian Inference Procedure}
This subsection describes the implementation of RoBART using $K$-fold cross-fitting, inspired by the double machine learning literature \citep{chernozhukov2018double}, in particular by the fold-specific bootstrap method in \cite{bach2024R}. In the context of difference-in-differences designs, \cite{BLY2024} develop a similar cross-fitted approach involving both prior and posterior adjustments.

Algorithm \ref{algorithm_cf} is the $K$-fold cross-fitting version of Algorithm \ref{algorithm}. We randomly partition the observation indices $\{1,\ldots,n\}$ into $K$ folds $(I_k)_{k=1}^K$, each of size $n/K$, assuming for notational simplicity that $n$ is divisible by $K$. For each fold $k$, we make draws of the fold-specific posterior for the conditional mean function and Bayesian-bootstrap weights. The fold-specific posterior of the conditional mean is obtained via Bayes' rule given the BART prior and fold-specific likelihood function. This differs from cross-fitting in the frequentist double machine learning literature, where the conditional mean function is estimated using observations outside the evaluation fold and evaluated on observations within that fold.
We make independent draws over different folds. Let $\widehat{\pi}^{(-k)}(\cdot)$ and $\widehat{m}^{(-k)}(\cdot)$ denote estimators of the propensity score $\pi_0(\cdot)$ and the conditional mean function $m_0(\cdot)$, respectively, constructed using all observations not belonging to fold $k$.

\begin{algorithm}[H]
    \caption{RoBART for $\mathbb{E}_0[Y_i]$ with $K$-fold cross-fitting}
    \label{algorithm_cf}

    (a) Take a $K$-fold random partition $(I_k)_{k=1}^K$ of the observation indices. The size of each fold $I_k$ is $n/K$.\;

    \For{$k=1,\ldots,K$}{

    (b) Use the subsample $\{(R_i,X_i^\top)\}_{i\notin I_k}$ to construct the estimator $\widehat{\pi}^{(-k)}(x)$. Set $\widehat{\gamma}^{(-k)}(r,x) =r/\widehat{\pi}^{(-k)}(x)$. \;

  (c) Use the subsample $\{(Y_i,X_i^\top):R_i=1\}_{i\notin I_k}$ to construct the estimator $\widehat{m}^{(-k)}(x)$. \;

  \For{$s=1,\ldots,S$}{
  (d) Make a draw from the posterior of $(m_\eta(X_i))_{i\in I_k}$ using the subsample 
$\{(Y_i,X_i^\top) : R_i=1\}_{i\in I_k}$, and denote it by $(m_\eta^{k,s}(X_i))_{i\in I_k}$.\;
            (e)  Generate $W_{i}^{k,s} = e_i^{k,s} / \sum_{j\in I_{k}} e_j^{k,s}$ with $e_i^{k,s} \stackrel{\text{iid}}{\sim} \mathrm{Exp}(1)$ for all $i\in I_{k}$.

            (f) Calculate $\check{\chi}_\eta^{k,s} =\chi_{\eta, \widehat \gamma}^{k,s}-\widehat{b}_\eta^{k,s}$, where
            \begin{equation}\label{recentered_bay_est}
  \chi_{\eta, \widehat \gamma}^{k,s}=\sum_{i\in I_k}W_i^{k,s}\left( m_\eta^{k,s}(X_i)
  +\widehat{\gamma}^{(-k)}(R_i,X_i)  \bigl( Y_i-m_\eta^{k,s}(X_i) \bigr)\right),
            \end{equation}
            and the debiasing term is
            \begin{equation}
                \label{debiased_bay_est}
                \widehat{b}_\eta^{k,s}= \frac{K}{n}\sum_{i\in I_k}\left(\widehat{\gamma}^{(-k)}(R_i,X_i)-1\right)\left( \widehat{m}^{(-k)}(X_i) - m_\eta^{k,s}(X_i)\right).
            \end{equation}
                    \vskip -.5cm
        }
    }

    \Output{
        $\big\{\check{\chi}_\eta^{s}=\frac{1}{K}\sum_{k=1}^K \check{\chi}_\eta^{k,s}:s=1,\ldots,S\big\}$
    }
\end{algorithm}

\section{BvM Theorems under High-level Assumptions}\label{sec:asympt}
Beyond the equivalence of point estimation, we aim to develop valid inference via showing new Bernstein--von Mises (BvM) theorems. Although our primary interest lies in establishing the BvM theorem for BART, we first develop the theory under high-level conditions before specializing the analysis. The robustness achieved through our additional debiasing step is therefore applicable to other types of priors beyond the BART framework. We begin with a version that relies on pilot estimators computed from an external sample in order to illustrate the main idea. This reduces the notational burden and highlights the core issue. In the next section, we prove the BvM theorem for the cross-fitted version. The proof essentially reduces to establishing the BvM theorem for the fold-specific posterior, which follows from the results developed in this section.

The one-step posterior of $\chi_{\eta, \widehat \gamma}$ is determined jointly by the posterior of the conditional mean and the Bayesian-bootstrap law. Recall that the conditional density associated with $m_\eta$ is denoted by $f_\eta$. For simplicity, we work with the conditional density class which only depends on the unknown conditional mean function. Such a class includes the standard Gaussian, Bernoulli, and Poisson outcomes, as discussed in \cite{BLY2022}. Accordingly, we denote it by $f_{m_\eta}(\cdot \mid x)$ hereafter. The posterior distribution is formally given by
\begin{align*}
	\Pi\big(m_{\eta}\in A, W^{(n)}\in B \bigm| Z^{(n)}\big)= \int_{B}\frac{\int_{A}\prod_{i=1}^{n}f^{R_i}_{m_\eta}(Y_i\mid X_i) \,\mathrm d\Pi_m(m_{\eta})}{\int \prod_{i=1}^{n}f^{R_i}_{m_\eta}(Y_i\mid X_i) \,\mathrm d\Pi_m(m_{\eta})}\,\mathrm d\Pi_W\big(W^{(n)}
	\bigm| Z^{(n)}\big).
\end{align*}
Regarding the centering point in the asymptotic normal approximation, we can consider any asymptotically efficient estimator $\widehat{\chi}$ with the following linear representation:
\begin{equation}\label{def:est:chi}
	\widehat{\chi}=\chi_0+\frac{1}{n}\sum_{i=1}^n \widetilde {\chi}_0(Z_i)+o_{P_0}(n^{-1/2}),
\end{equation}
where $\widetilde{\chi}_0=\widetilde{\chi}_{\eta_0}$ is the	 efficient influence function given in \eqref{eif} under $\eta=\eta_0$. Denote its variance by $\textsc v_0=\E_0\left[ \widetilde{\chi}_0^2(Z)\right]$. We write $\mathcal{L}_{\Pi}(\sqrt{n}(\chi_\eta-\widehat{\chi})\mid Z^{(n)})$ for the posterior distribution of $\sqrt{n}(\chi_\eta-\widehat{\chi})$.

Below, we consider some measurable sets $\mathcal H^m_n$ of functions $m_{\eta}$ such that $\Pi(m_{\eta}\in\mathcal{H}^m_n\mid Z^{(n)})\to_{P_0} 1$. With a slight abuse of notation, we also denote $\mathcal{H}_n=\{\eta:m_{\eta}\in\mathcal{H}_n^m\}$ where we index the conditional mean function $m_{\eta}$ by its subscript $\eta$. We introduce the notation $\|\phi\|_{ 2,\mu_0}:= \sqrt{\int \phi^2(x)\,\mathrm d\mu_0(x)}$ for all $\phi\in L^2(\mu_0):=\{\phi:\|\phi\|_{ 2,\mu_0}<\infty\}$, as well as the supremum norm $\|\cdot\|_\infty$.
We introduce the notation for the conditional variance $\sigma_0^2(x)= \mathbb E_0[(RY-m_0(X))^2\mid R=1,X=x]$ and assume throughout the paper that $\sigma_0\in  L^2(\mu_0)$.
We denote the support of a random variable $V$ by $\mathcal V$.

\begin{assumption}[Identification]\label{Assump:ID}
	We observe an i.i.d. sample $\{(R_iY_i,R_i,X_i^\top)^\top\}_{i=1}^n$,  where the propensity score $\pi_0$ satisfies $\pi_0(x)=\P(R_i=1\mid Y_i=y, X_i=x)$ for all $(y,x)\in\mathcal Y\times \mathcal X$ and $\inf_{x\in\mathcal{X}}\pi_0(x)\geq c$ for some constant $c>0$. 
\end{assumption}
Assumption \ref{Assump:ID} imposes a missing-at-random (MAR) assumption and an overlap assumption, both of which are sufficient for identifying the conditional mean function $m_0(x)$.

	\begin{assumption}[Rates of Convergence]\label{Assump:Rate}
The pilot estimators $\widehat \pi$ and $\widehat m$, which are based on an auxiliary sample independent of $Z^{(n)}$, satisfy 	$\Vert \widehat{\pi}-\pi_0\Vert_{2, \mu_0}=O_{P_0}(r_n) $ and $\Vert 1/\widehat\pi\Vert_{\infty}=O_{P_0}(1)$. In addition, 
		\begin{equation*}
			\Vert \widehat{m}-m_0\Vert_{ 2,\mu_0}=O_{P_0}(\varepsilon_n)~\text{ and }~\sup_{\eta\in\mathcal{H}_n}\Vert m_\eta-m_0\Vert_{2,\mu_0}\lesssim \varepsilon_n,
		\end{equation*}
	for some measurable sets $\mathcal{H}_n=\{\eta:m_{\eta}\in\mathcal{H}_n^m\}$ such that $\Pi(m_{\eta}\in\mathcal{H}^m_n\mid Z^{(n)})\to_{P_0} 1$,	where $\max\{\varepsilon_n, r_n\}\to 0$ and $\sqrt{n}\,\varepsilon_nr_n\to 0$. 
	\end{assumption}
	 Assumption \ref{Assump:Rate} imposes sufficiently fast convergence rates of the estimators of the conditional mean function $m_0$ and the propensity score $\pi_0$. In practice, one can explore the recent proposals from \cite{ChernozhukovNeweySingh2020b} and \cite{hirshberg2021augmented}. The posterior contraction rate for the conditional mean can be derived by verifying high-level assumptions of \cite{ghosal2000rates}.
	 
	\begin{assumption}[Complexity]\label{Assump:Donsker}
(i)	For $\mathcal{G}_n=\{m_{\eta}(\cdot):\eta\in \mathcal{H}_n\}$ it holds that
\begin{equation}\label{GlivenkoCantelli}
\sup _{m_{\eta} \in \mathcal{G}_n}\left|(\mathbb{P}_n-P_0) m_{\eta}\right| =o_{P_0}(1).
\end{equation}
	 Let $G_n$ be the envelope function of the function class $\mathcal{G}_n$ satisfying
	\begin{equation}\label{MomentConditions}
		\lim_{C\to\infty}\limsup_{n\to\infty}\mathbb{E}_0[G_n^2\mathbbm1\{G_n>C\}]=0,~~~	\mathbb{E}_0[G_n^4]=o(n).
	\end{equation}
(ii) Furthermore, we assume that
\begin{equation}\label{ProductSE}
\sup_{\eta\in\mathcal{H}_n}\left|\mathbb{G}_n\left[\left(\gamma_0-\widehat \gamma\right)(m_\eta-m_0)\right]\right|=o_{P_0}(1).
\end{equation}		
	\end{assumption}

 Assumption \ref{Assump:Donsker} (i) restricts the function class $\mathcal{G}_n$ to form a $P_0$-Glivenko--Cantelli class; see Section 2.4 of \cite{van1996empirical}. The moment conditions on the envelope functions follow \cite{yiu2023corrected} and address the uniformity issues when showing the convergence of the conditional Laplace transform. Assumption \ref{Assump:Donsker} (ii) imposes a new stochastic equicontinuity condition on the product structure involving $\widehat\gamma$ and $m_\eta$, which sets us apart from the existing literature. In contrast to this assumption, \cite{ray2020causal} and \cite{yiu2023corrected} both require a stochastic equicontinuity condition $\sup_{\eta\in\mathcal{H}^m_n}\left|\mathbb{G}_n\left[m_\eta-m_0\right]\right|=o_{P_0}(1)$\footnote{Because a nonparametric prior is also assigned to the propensity score, \cite{yiu2023corrected} further require the propensity score function to belong to a Donsker class.}. As demonstrated by \cite{BLY2022}, this is weaker than directly imposing the Donsker property on $(m_{\eta}-m_0)$. For the H\"older smooth class, \cite{BLY2022} show that the low-order differentiability of the conditional mean can be compensated by exploiting the high-order smoothness of the propensity score, and vice versa.  
Note that $\sup_{\eta\in\mathcal{H}^m_n}\left|\mathbb{G}_n\left[m_\eta-m_0\right]\right|$ diverges for the non-Donsker class, by Sudakov's inequality \citep{van1996empirical}.\\

We now present a semiparametric Bernstein–von Mises theorem, which establishes asymptotic normality of
the posterior distribution, modulo a bias term. This asymptotic equivalence result is established using the so-called \textit{bounded Lipschitz distance} $d_{BL}$ on probability distributions on $\mathbb R$; see Chapter 11 of \cite{dudley2002}. 
 
\begin{theorem}\label{thm:BvM} Let Assumptions \ref{Assump:ID}--\ref{Assump:Donsker} hold.
	Then, for the baseline one-step posterior, we have	
		\begin{equation*}
		d_{BL}\left(\mathcal{L}_{\Pi}(\sqrt{n}(\chi_{\eta, \widehat \gamma}-\widehat{\chi}-b_{0,\eta})\mid Z^{(n)}), N(0,\textsc v_0) \right)\to_{P_0} 0,
	\end{equation*}
where 	$b_{0,\eta}=	\mathbb{P}_n[(\gamma_0-1)(m_0-m_{\eta})]$.
	\end{theorem}
The two parts of Assumption \ref{Assump:Donsker} play distinct roles in the proof. Part (i) delivers the conditional convergence of the Laplace transform of the Bayesian-bootstrap process through the $P_0$-Glivenko--Cantelli property of $\mathcal{G}_n$, while part (ii) controls the remainder term $\mathbb{P}_n[(\gamma_0-\widehat{\gamma})(m_{\eta}-m_0)]$ arising from the pilot estimation of the propensity score.

	The previous result shows that, under our double-robust smoothness conditions, the baseline one-step posterior satisfies the BvM theorem only up to a bias term. Specifically, the bias term is given by
          \begin{align*}
          b_{0,\eta} 
          &= \mathbb{P}_n[m_{\eta}(X) +\gamma_0 (Y-m_{\eta})]-\mathbb{P}_n[m_0 +\gamma_0 (Y-m_0)]\\
           &= \mathbb{P}_n\left[m_{\eta}(X) +\frac{R}{\pi_0(X)} (Y-m_{\eta}(X))\right]-\mathbb{P}_n\left[m_0(X) +\frac{R}{\pi_0(X)} (Y-m_0(X))\right],
          \end{align*}
        which measures the empirical influence-function discrepancy between $m_{\eta}$ and $m_0$. 
      The difference from frequentist doubly robust estimation arises because $m_{\eta}$ is a random function under the nonparametric Bayesian posterior. As a consequence, the resulting bias term need not be asymptotically negligible. In general, it vanishes asymptotically only under Donsker conditions on the conditional mean function; see Remark \ref{rmk:donsker}.
\begin{remark}[Bias Equivalence for Prior Correction]\label{rem:bias:equivalence}
Under the product-rate condition imposed in Assumption \ref{Assump:Rate}, Theorem 3.1 of \cite{BLY2022} establishes a semiparametric BvM theorem for the prior-adjustment approach of \cite{ray2020causal}, denoted by $\chi_\eta^{PA}$, with bias term $b_{0,\eta}$. Combined with Theorem \ref{thm:BvM} above, this implies that the prior-adjusted posterior and the one-step posterior considered here exhibit the same first-order bias. In particular, by the triangle inequality for the bounded Lipschitz distance, we have the asymptotic equivalence
 \begin{equation*}
	d_{BL}\left(\mathcal{L}_{\Pi}(\sqrt{n}(\chi^{PA}_\eta-\widehat{\chi}-b_{0,\eta})\mid Z^{(n)}), \mathcal{L}_{\Pi}(\sqrt{n}(\chi_{\eta, \widehat \gamma}-\widehat{\chi}-b_{0,\eta})\mid Z^{(n)}) \right)\to_{P_0} 0.
\end{equation*}
In other words, the one-step posterior exhibits exactly the same bias term as the plug-in version using the adjusted prior of \cite{ray2020causal}.
\end{remark}	
	
We next relax the Donsker condition by allowing for pilot estimators of the unknown functional parameters in $b_{0,\eta}$, while retaining the weaker product-rate condition in \eqref{ProductSE}.
We explicitly correct the posterior distribution, following our proposed methodology in Algorithm \ref{algorithm}. Recall the definition of the bias correction term  $\widehat{b}_{\eta}$ given  in Algorithm \ref{algorithm}:
\begin{align}\label{recentering_term}
\widehat{b}_{\eta}=	\mathbb{P}_n[(\widehat \gamma-1)(\widehat m-m_{\eta})].
\end{align}
 Under our conditions, $\widehat{b}_{\eta}$ consistently estimates the posterior-specific bias $b_{0,\eta}$ uniformly over the relevant posterior set, so that subtracting $\widehat{b}_{\eta}$ removes the first-order centering error. In the present one-step construction, this correction also yields the exact AIPW representation discussed in Section \ref{sec:model}.

\begin{theorem}\label{coro:BvM:debias}
	Let Assumptions \ref{Assump:ID}--\ref{Assump:Donsker} hold. Then, we have
	\begin{equation*}
		d_{BL}\left(\mathcal{L}_{\Pi}(\sqrt{n}(\check\chi_\eta-\widehat{\chi})\mid Z^{(n)}),  N(0,\textsc v_0)  \right)\to_{P_0} 0.
	\end{equation*}
\end{theorem}

An important consequence of the BvM results is the frequentist validity of the Bayesian credible set built from the corrected posterior. For any $\alpha\in(0,1)$, the Bayesian credible set $\mathcal{C}_n(\alpha)$ for the corrected draw $\check\chi_\eta=\chi_{\eta, \widehat \gamma}-\widehat b_{\eta}$ satisfies $	P_0\big(\chi_0\in  \mathcal{C}_n(\alpha)\big) \to 1-\alpha.$

Below,  let the AIPW estimator by \cite{robins1994regression} be  $\widehat\chi_{\mathrm{AIPW}}:=\mathbb P_n[\widehat m(X)+\widehat\gamma(R,X)(Y-\widehat m(X))]$. Under our assumptions, it is straightforward to see that $\widehat{\chi}_{\mathrm{AIPW}}$ satisfies the asymptotic linear representation in \eqref{def:est:chi} and hence achieves the semiparametric efficiency bound. Together with the discussion in Section \ref{sec:mot}, this yields the following result.
\begin{corollary}\label{coro:BvM:debias:AIPW}
	Let Assumptions \ref{Assump:ID}--\ref{Assump:Donsker} hold. Then, we have
\begin{align}
\check\chi_\eta=\widehat\chi_{\mathrm{AIPW}}+(\mathbb P_n^*-\mathbb P_n)[ m_\eta(X)+\widehat\gamma(R,X)(Y- m_\eta(X))]\label{EquivalenceID}
\end{align}
and
	\begin{equation*}
		d_{BL}\left(\mathcal{L}_{\Pi}(\sqrt{n}(\check\chi_\eta-\widehat\chi_{\mathrm{AIPW}})\mid Z^{(n)}),  N(0,\textsc v_0)  \right)\to_{P_0} 0.
	\end{equation*}
\end{corollary}
This corollary provides a direct link between the corrected posterior and conventional semiparametric estimation. Its finite-sample center is exactly the AIPW estimator based on the pilot nuisance estimates, while its asymptotic distribution is Gaussian with the semiparametric efficiency bound. Thus, the correction does not alter the first-order efficient point estimator, but instead constructs a posterior distribution around this center whose uncertainty is generated jointly by the posterior variation in the conditional mean function and the Bayesian bootstrap. This distinction will also be useful for interpreting the finite-sample comparison with DML in Section \ref{sec:numerical}.

For intuition, note that the difference $\check\chi_\eta-\widehat\chi_{\mathrm{AIPW}}$ in (\ref{EquivalenceID}) resembles a centered bootstrap version of the AIPW estimator. Hence, conditional weak convergence to the desired normal limiting distribution is to be expected. That said, the technical challenge we address concerns uncertainty arising from the nonparametric Bayesian modeling of $m_{\eta}$, which differs substantially from the semiparametric bootstrap theory underlying existing frequentist constructions.

If the Donsker property holds, our procedure remains valid even when the pilot estimators are computed over the same sample. This can be easily seen as the debiasing term becomes asymptotically negligible. In other related causal inference problems, we observe that the incorporation of such a debiasing term can improve the finite-sample performance even without any sample splitting \citep{BLY2022,BLY2024}.

\section{A BvM Theorem for the Cross-fitted Posterior under Primitive Conditions}\label{sec:bart}
Now we present the theoretical justification for the $K$-fold cross-fitted posterior under primitive conditions tailored to BART priors. The key ingredient of our construction involves the fold-specific posterior for the conditional mean function. 
\subsection{Tree Constructions}
We now provide a detailed specification of the BART-type priors. Because the construction applies to any fold-specific posterior, we drop the additional subscript or superscript on $k$ to lower the notational burden. 

The BART prior expresses the unknown function as a sum of many binary regression trees. Each individual tree consists of a set of internal decision nodes which define a partition of the covariate space, as well as a set of terminal nodes or leaves. Those partitions are generated by recursively applying some binary split rules of the covariate space of the form $\{x_j \leq \tau \}$ versus $\{x_j>\tau \}$ where $x_j$ signifies that the $j$-th coordinate is chosen for the splitting and $\tau$ is the split point. The good approximation property of tree-based learners relies on finding a fine partition scheme that divides the data into more homogeneous groups and learning a piecewise-constant function on the partition. For this purpose, we introduce the notion of the split-net from \cite{rockova2020posterior}. Given an increasing integer sequence $b_n$, a split-net $\mathcal{X}=\{x_i=(x_{i1},\ldots,x_{ip})^\top\in[0,1]^p,i=1,\ldots,b_n \}$ is a discrete collection of $b_n$ split-candidates $x_i$ at which possible splits occur along coordinates. For a set $\Omega\subset\mathbb{R}^p$, we denote the $j$-th projection mapping of $\Omega$ by $[\Omega]_j=\{x_j\in \mathbb{R}:(x_1,\ldots,x_p)^\top\in \Omega \}$.

The construction of each individual tree starts with some root node in $[0,1]^p$ at depth $\delta=0$, where the depth of a node means the number of nodes along the path from the root node down to that node. Any binary tree can be characterized by (i) the probability that a node at depth $\delta$ is nonterminal, (ii) the distribution on the splitting variable assignments at each splitting node, and (iii) the rules with which the splits are made. Referring to point (i), each node at depth $\delta\in\{0,1,2,\ldots\}$ is split (hence nonterminal) with some prior probability depending on the depth. When it comes to point (ii), we follow \cite{linero2018dart} and \cite{linero2018bart} in using a sparse Dirichlet prior. This approach chooses a splitting covariate $j$ from a proportion vector $\vartheta=(\vartheta_1,\ldots,\vartheta_p)^\top$ belonging to the $p$-dimensional simplex. Finally, for point (iii), if a node corresponding to a box $\Omega$ is split, a splitting coordinate is drawn from the proportion vector and a split point $\tau_j$ is chosen randomly from $[\mathcal{X}]_j\cap\textsf{int}([\Omega]_j)$ for a given split-net $\mathcal{X}$. The procedure continues until all nodes are terminal. We further denote the cardinality of the split points in the split-net $\mathcal{X}$ projected onto the $j$-th coordinate by $b_j(\mathcal{X})$ for $j=1,\ldots,p$.

We pick a fixed number of trees $T$ and assume an independent product prior for the tree ensemble, $\Pi(\mathcal{E})=\prod_{t=1}^T \Pi(\mathcal{T}^t)$. \cite{chipman2010bart} recommend taking $T=200$ based on extensive numerical evidence. For a given split-net $\mathcal{X}$ and for each $1\leq t\leq T$, we denote by $\mathcal{T}^t=(\Omega^t_1,\ldots,\Omega^t_{L^t})$ an $\mathcal{X}$-tree partition of size $L^t$, with step heights $\bm{\beta}^t=(\beta_1^t,\ldots,\beta_{L^t}^t)\in\mathbb{R}^{L^t}$. An additive tree-based learner is fully described by a tree ensemble $\mathcal{E}=\{\mathcal{T}^1,\ldots,\mathcal{T}^T \}$ and terminal node parameters $\mathcal{B}=(\bm{\beta}^{1},\ldots,\bm{\beta}^{T})^\top\in\mathbb{R}^{\sum_{t=1}^T L^t}$ as follows
\begin{equation}\label{TreeFunc}
	m_{\mathcal{E},\mathcal{B}}(x)=\sum_{t=1}^T\sum_{l=1}^{L^t}\beta_l^t\mathbbm1\{x\in \Omega^t_l\}.
\end{equation}
Given an ensemble $\mathcal{E}$ of trees, we denote $\mathcal{M}_{\mathcal{E}}:=\{	m_{\mathcal{E},\mathcal{B}}(x):\mathcal{B}\in\mathbb{R}^{\sum_{t=1}^TL^t} \}$. If $\mathcal{E}$ consists of a single tree $\mathcal{T}$, we denote  $\mathcal{M}_{\mathcal{E}}$ by $\mathcal{M}_{\mathcal{T}}$.  

\subsection{Primitive Conditions}
Now we introduce primitive conditions about the conditional mean function. Binary decision trees or forests offer good approximation properties for the H\"older \textit{continuous class}, which is indexed by an exponent $\alpha\in(0,1]$. This parameter $\alpha$ does not exceed one, which is standard in the literature studying the piecewise-constant estimators or priors.  Below we introduce the norm  $\Vert \phi\Vert_{\mathcal{H}^{\alpha}}:= \sup_{x,y\in[0,1]^p}|\phi(x)-\phi(y)|/\Vert x-y\Vert_2^{\alpha}$, for functions $\phi:[0,1]^p\to \mathbb{R}$, with some $\alpha>0$, where $\|\cdot\|_2$ denotes the Euclidean norm. The true conditional mean function is assumed to belong to the following class of functions. 
\begin{definition}
We denote the space of uniformly $\alpha$-H\"older \textit{continuous} functions that only depend on some subset of covariates $\mathcal{S}$ as follows,
	\begin{align*}
		\mathcal{F}_p(\alpha,\mathcal{S}):=&\Big\{m:[0,1]^p\to \mathbb{R}: \Vert m\Vert_{\mathcal{H}^{\alpha}}<\infty \text{ and }m~ \text{is constant in the directions}~\{1,\ldots,p \}\setminus \mathcal{S} \Big\}
	\end{align*}
	where $\alpha\in (0,1]$ and $\Vert m\Vert_{\mathcal{H}^{\alpha}}$ is the H\"older norm. 
\end{definition}
BART can also exploit sparsity in the relevant regressors and perform variable selection. For the set $\mathcal{S}_0$ with its cardinality $|\mathcal{S}_0|=q_0<p$, we assume $m_0\in 	\mathcal{F}_p(\alpha,\mathcal{S}_0)$. In addition, we also explore the case where $m_0$ is additively separable in terms of low-dimensional covariates. Consider the following additive class with $T_0$ components
 \begin{equation*}
 	\mathcal{F}^{add}_p(\bm{\alpha},\bm{\mathcal{S}}_0):=\left\{m_0(x)=\sum_{t=1}^{T_0}m_0^t(x),~~\text{such that}~~m_0^t\in \mathcal{F}_p(\alpha^t,\mathcal{S}^t_0)\right\},
 \end{equation*}
where $\bm{\alpha}:=(\alpha^t)_{t=1}^{T_0}$ and $\bm{\mathcal{S}}_0:=(\mathcal{S}^t_0)_{t=1}^{T_0}$. We denote $q_0^t=|\mathcal{S}^t_0|$ for the subset $\mathcal{S}^t_0\subset\{1,\ldots,p\}$. We define
\begin{equation*}
	\varepsilon_{n}:=n^{-\alpha/(2\alpha+q_0)}\sqrt{\log n},
\end{equation*}
related to the rate of posterior contraction for $m_0\in 	\mathcal{F}_p(\alpha,\mathcal{S}_0)$. When considering the additive case $m_0\in 	\mathcal{F}^{add}_p(\bm{\alpha},\bm{\mathcal{S}}_0)$, we define $\varepsilon^{add}_{n}=\sqrt{\sum_{t=1}^{T_0}(\varepsilon^t_{n})^2}$, where $\varepsilon^t_{n}:=n^{-\alpha^t/(2\alpha^t+q^t_0)}\sqrt{\log n}$.

The case where $m_0\in \mathcal{F}_p(\alpha,\mathcal{S}_0)$ provides a clean illustration of our theory, which shows that our additional debiasing step is essential if $q_0>1$. The class $\mathcal{F}_p(\alpha,\mathcal{S}_0)$ fails to satisfy the Donsker property whenever $\alpha<q_0/2$. A subtle consequence is that the size of the resulting sieve set used to approximate the H\"older continuous function becomes too large, so that the standard maximal inequality via the entropy bound fails to deliver the stochastic equicontinuity. The same issue occurs for the additive class whenever $\alpha^t<q^t_0/2$ for any $1\leq t\leq T_0$.

Next, we list assumptions on the outcome model, the split-net, and the prior, followed by a remark.
\begin{assumption}[Model Specification]\label{Assump:BARTModel}
 Under $P_0$, the conditional density of $Y$ given $(R=1,X=x)$ is Gaussian with unit variance, i.e.,
\begin{equation*}
	f_0(y\mid x)=\frac{1}{\sqrt{2\pi}}\exp\left(-\frac{(y-m_{0}(x))^2}{2}\right).
\end{equation*}
Moreover, let $\log p\lesssim n^{q_0/(2\alpha+q_0)}$ if $m_0\in 	\mathcal{F}_p(\alpha,\mathcal{S}_0)$, and $\log p\lesssim \min_{1\leq t\leq T_0}n^{q^t_0/(2\alpha^t+q^t_0)}$ if $m_0\in \mathcal{F}^{add}_p(\bm{\alpha},\bm{\mathcal{S}}_0)$.
\end{assumption}
\begin{assumption}[Tree-based Partition]\label{Assump:BARTTree}
	 The split-net $\mathcal X$ satisfies the following conditions: (i) $\max_{1\leq j\leq p}\log b_j(\mathcal{X})\lesssim \log n$.
	 (ii) One can construct an $\mathcal{X}$-tree partition $\widehat{\mathcal{T}}$ such that there exists $m_{0,\widehat{\mathcal T},\widehat{\bm{\beta}}}$ with its step heights $\widehat{\bm{\beta}}$, satisfying $\Vert m_0-m_{0,\widehat{\mathcal T},\widehat{\bm{\beta}}}\Vert_\infty\lesssim\varepsilon_{n}$ if $m_0\in 	\mathcal{F}_p(\alpha,\mathcal{S}_0)$. For $m_0\in \mathcal{F}^{add}_p(\bm{\alpha},\bm{\mathcal{S}}_0)$, we assume this tree-based approximation exists for each individual component in the additive function, that is, $\Vert m^t_0-m_{0,\widehat{\mathcal T}^t,\widehat{\bm{\beta}}^t}\Vert_\infty\lesssim\varepsilon^t_{n}$ for $t=1,\ldots,T_0$.
\end{assumption}
\begin{assumption}[Prior]\label{Assump:BARTPrior}
(i) For a fixed integer $T\geq 1$, each tree $\mathcal{T}^t$, $t=1,\ldots,T$, is independently assigned a tree prior with the following Dirichlet sparsity, that is, the $j$-th covariate is chosen for splitting the nonterminal nodes from a proportion vector $\vartheta:=(\vartheta_1,\ldots,\vartheta_p)^\top$, such that $(\vartheta_1,\ldots,\vartheta_p)^\top\sim \textrm{Dir}(\zeta/p^{\xi},\ldots,\zeta/p^{\xi})$ with $\zeta>0$ and $\xi>1$. (ii) Given any tree, each node at depth $\delta\in\{0,1,2,\ldots\}$ is split with some prior probability $\nu^{\delta+1}$ for some $\nu\in (0,1/2)$. (iii) Given $L^1,\ldots,L^T$ induced by $\mathcal{E}$, we consider the independent priors for the step heights:
\begin{equation*}
	\textrm d\Pi(\mathcal{B}\mid L^1,\ldots,L^T)=\prod_{t=1}^T\prod_{l=1}^{L^t}\phi_T(\beta_l^t)\,\textrm d\beta_l^t,
\end{equation*}
where $\phi_T(\cdot)$ is some bounded and compactly supported probability density function that can depend on the number of trees $T$. 
\end{assumption}
\begin{remark}[Discussion of Assumptions]
The Gaussian likelihood with  unit variance in Assumption \ref{Assump:BARTModel} is standard in the literature. One can also incorporate the unknown variance term and assign a prior as in \cite{xie2020adapt} and \cite{jeong2023art}. The asymptotic analysis allows for a large ambient dimension $p$. The requirement about the growth of $\log p$ simplifies the presentation of the contraction rate. Otherwise, one needs to incorporate an additional term such as $\sqrt{q_0(\log p)/n}$ in the high-dimensional setup. The construction of the piecewise-constant approximation with the tree partition, which satisfies Assumption \ref{Assump:BARTTree}, can be found in Section 4 of \cite{jeong2023art}. 

Assumption \ref{Assump:BARTPrior} lists the specification of priors in BART. The control of the unique elements in the split-net along each coordinate is crucial to establish the posterior rate of contraction. The original algorithm in \cite{chipman2010bart} does not account for sparsity of the true regression model. \cite{linero2018dart} suggests the sparse Dirichlet prior on splitting coordinates to perform variable selection.
Following \cite{rockova2020theory}, the splitting probabilities decay exponentially with respect to the depth $\delta$, which gives rise to the desired exponential tail of the total tree sizes. The compact support condition on each step height is assumed for technical reasons when verifying high-level conditions in \cite{ghosal2000rates}; see \cite{xie2020adapt} and \cite{jeong2023art}.  
\end{remark}
\subsection{A New BvM Theorem for BART}
In our Algorithm \ref{algorithm_cf}, when we sample from the posterior distributions of $\chi_{\eta, \widehat \gamma}^{k}$ for $k=1,\ldots,K$, the draws from the posterior distribution of the conditional mean function $m^k_{\eta}$ and the Bayesian-bootstrap weights $W^{(n_k)}$ are generated independently across folds. As such, the \textit{posterior} law under consideration in our algorithm can be written as the convolution of fold-specific posterior distributions. For a pair of independent random variables $U_1$ and $U_2$ with marginal laws $\mathcal{U}_1$ and $\mathcal{U}_2$, respectively, we denote their convolution by $\mathcal{L}_{\mathcal{U}_1}(U_1)\ast \mathcal{L}_{\mathcal{U}_2}(U_2)$, which represents the law of their sum.

Recall from Section \ref{sec:model} that $K$ divides $n$, so that every fold is of the same size, i.e., $|I_1|=\cdots=|I_K|=n/K$. We define the DML estimator by $\widehat{\chi}_{\mathrm{DML}}=K^{-1}\sum_{k=1}^K \widehat{\chi}^k$ where in this section:
\begin{align*}
	\widehat{\chi}^k:=	\frac{K}{n}\sum_{i\in I_k}\left(\widehat{m}^{(-k)}(X_i)+\widehat{\gamma}^{(-k)}(R_i,X_i)\left(Y_i-\widehat{m}^{(-k)}(X_i)\right)\right).
\end{align*}
For each fold $k$, our corrected posterior estimator is given by $\check\chi_\eta^k=\chi_{\eta, \widehat \gamma}^k-\widehat b_\eta^k$. As a point estimator, we use the average of the posterior means of the fold-specific corrected estimators. Formally,
\begin{equation}
\frac{1}{K}\sum_{k=1}^K\mathbb{E}_{\Pi}\left[\check\chi_\eta^k\mid Z^{(n_k)}\right]
=\frac{1}{K}\sum_{k=1}^K\widehat{\chi}^k,
\end{equation}
where the equality follows from the same argument as in \eqref{EquivalenceAIPW}, since the pilot estimators used for fold $k$ are computed using observations outside $I_k$.

The numerical equivalence of the posterior mean and the DML estimator is agnostic to the type of nonparametric priors for the conditional mean, as well as the pilot estimators $(\widehat{m},\widehat{\pi})$. Nonetheless, our simulation results favor BART-type modeling and reusing its posterior mean as the pilot estimator $\widehat{m}^{(-k)}$. This is also close to the findings of \cite{dorie2019auto}, where targeted MLE (TMLE) with BART point estimators is among the top performers in their large-scale data competition. 

 Our main theoretical contribution is that, besides the point estimator, the credible sets constructed from the fold-specific posteriors are valid. Next, we present a formal statement of the validity of the $K$-fold posterior cross-fitting procedure under primitive conditions for the BART priors. Regarding the convolution in the stated theorem, we do not claim conditional independence of the fold-specific posteriors under the ground-truth probability law $P_0$, conditional on the entire observed data $Z^{(n)}$. Rather, we deliberately generate posterior draws of the conditional mean function and the bootstrap weights separately across different folds. After applying Bayes' rule to those components, the posterior laws $\mathcal{L}_{\Pi}(\cdot\mid Z^{(n_k)})$, which encode the randomness of the conditional mean function and the bootstrap weights, are made independent over different folds. 
\begin{theorem}\label{thm:BvM:BART}
Let Assumptions \ref{Assump:ID} and  \ref{Assump:BARTModel}--\ref{Assump:BARTPrior}  hold. In addition, we assume that $\|1/\widehat \pi^{(-k)}\|_\infty=O_{P_0}(1)$ for $1\leq k\leq K$. If $m_0\in 	\mathcal{F}_p(\alpha,\mathcal{S}_0)$, $\Vert \widehat{m}^{(-k)}-m_0\Vert_{ 2,\mu_0}=O_{P_0}(\varepsilon_n)$, and $\sqrt{n}\,\varepsilon_{n}\Vert \widehat{\pi}^{(-k)}-\pi_0\Vert_{\infty}\to_{P_0}0$, 
then
\begin{equation*}
		d_{BL}\left(\mathcal{L}_{\Pi}\left(\frac{\sqrt{n}}{K}(\check\chi^1_\eta-\widehat{\chi}^1)\mid Z^{(n_1)}\right)\ast\ldots\ast\mathcal{L}_{\Pi}\left(\frac{\sqrt{n}}{K}(\check\chi^K_\eta-\widehat{\chi}^K)\mid Z^{(n_K)}\right), N(0,\textsc v_0) \right)\to_{P_0} 0.
	\end{equation*}
The same conclusion holds if $m_0\in \mathcal{F}^{add}_p(\bm{\alpha},\bm{\mathcal{S}}_0)$, $\Vert \widehat{m}^{(-k)}-m_0\Vert_{2,\mu_0}=O_{P_0}(\varepsilon^{add}_{n})$ and $\sqrt{n}\,\varepsilon^{add}_{n}\Vert \widehat{\pi}^{(-k)}-\pi_0\Vert_{\infty}\to_{P_0}0$.
\end{theorem}

Theorem \ref{thm:BvM:BART} establishes a BvM result for BART without requiring the Donsker property for either the propensity score or the conditional mean function. We first argue that it suffices to consider each fold-specific posterior separately. Then, the desired result follows from verifying the high-level assumptions from the previous section. 

\begin{remark}[Donsker Class]\label{rmk:donsker}
One can establish the BvM theorem
under the Donsker property of $\mathcal{H}_n$, so that $b_{0,\eta}$ becomes asymptotically negligible itself. This condition holds if $\alpha>q_0/2$ for $m_0\in 	\mathcal{F}_p(\alpha,\mathcal{S}_0)$. With the additive structure, the above property holds for $m_0\in \mathcal{F}^{add}_p(\bm{\alpha},\bm{\mathcal{S}}_0)$ if $\alpha^t> q^t_0/2$ for all $1\leq t\leq T_0$. 
For semiparametric Bayesian inference, such Donsker properties are imposed in Assumption 2(c) in \cite{yiu2023corrected} or Condition (3.12) in \cite{ray2020causal}.
Our results exploit the regularity of the propensity score, expressed in terms of a new stochastic equicontinuity condition, to restore the frequentist validity of our robust procedure. 
\end{remark}

\begin{remark}[Smoothed BART]\label{rmk:smoothedbart}
Given the nonsmooth feature of BART, it is natural to consider the smoothed BART (see \cite{linero2018bart}), because it exploits the smoothness of the conditional mean. The same trade-off between the smoothness of the conditional mean and propensity score continues to hold when the additional bias correction is applied. Consider functions in the H\"older \textit{smooth} class, i.e., the space of functions on $[0,1]^p$ with bounded partial derivatives up to order $\lfloor \alpha\rfloor$, where $\lfloor \alpha\rfloor$ is the largest integer strictly less than $\alpha$ and such that the partial derivatives of order $\lfloor \alpha\rfloor$ are H\"older \textit{continuous} of order $\alpha-\lfloor \alpha\rfloor$. If the true function only depends on at most $q_0$ covariates, Theorem 2 in \cite{linero2018bart} states that the posterior rate of contraction is of the order $O(n^{-\alpha/(2\alpha+q_0)}\log (n))$. If the conditional mean function and propensity score function are in such H\"older smooth classes with smoothness indices $(\alpha_m,\alpha_{\pi})$ and they only depend on $q_0$ regressors, our robust approach is asymptotically normal when $\sqrt{\alpha_m \alpha_{\pi}}>q_0/2$. In comparison, the Donsker class restriction will force $\min\{\alpha_m, \alpha_{\pi}\}>q_0/2$. 
\end{remark}

\section{Numerical Studies}\label{sec:numerical}
In this section, we first examine the finite-sample performance of the proposed robust BART procedure for the mean response $\chi_0=\mathbb{E}_0[Y_i]$ in missing-data models. Then, building on Appendix \ref{appendix:other_para}, which extends the robust BART to the average treatment effect (ATE) framework, we apply the method to the well-known National Health and Nutrition Examination Survey data and revisit the average effect of participation in meal programs on students’ body mass index.

\subsection{Monte Carlo Simulation} \label{sec:simulation}
The data-generating process for i.i.d. observations is specified as follows. For each unit $i=1,\ldots, n$, we generate the observed variables $(R_iY_{i}, R_i, X_{i}^\top)$ by $X_i = (X_{i1},\dots, X_{i5})^\top$ where $X_{i1}, X_{i2}, X_{i3}\sim N(0,1)$, $X_{i4}\sim \text{Bernoulli}(0.5)$, and $X_{i5}$ is a categorical variable taking values $\{1, 2, 3\}$ with equal probability. The distributions of $R_i$ and $Y_i$ are given by
\begin{eqnarray*}
R_i\mid X_i \sim  \text{Bernoulli}\left(\Phi\left[e(X_{i})\right]\right),\quad
Y_i \mid X_i  \sim  N\left(m(X_{i}), 1\right), 
\end{eqnarray*}
where $\Phi$ denotes the cumulative distribution function of the standard normal distribution. We analyze the finite-sample effects of varying the complexity of the conditional mean function $m$ and the propensity score for sample sizes $n \in \{1000, 2000\}$. Throughout our simulations, the number of Monte Carlo replications
is set to 1,000.

We consider four designs for the function $e(\cdot)$ that determines the propensity score by $\pi(x)=\Phi(e(x))$ and for the conditional mean function $m(\cdot)$. In all four designs, neither $e(\cdot)$ nor $m(\cdot)$ is additively separable in $x$.  The complexity of the  nuisance functions increases as more interaction terms are added.
\begin{itemize}
\item[] Design I: \hspace*{.28cm}$e(x)= -0.2x_1 + 0.4x_1x_3$,  $m(x)=1+ x_1x_2 + x_1x_3 + x_2 + x_4 + h(x_5)$,
\item[] Design II: \hspace*{.135cm}$e(x)= -0.2x_1 + 0.4x_1x_3$, $m(x) = 1+ x_1x_3 + x_2x_3 + x_2x_4 + h(x_5)$,
\item[] Design III: $e(x)= -0.2x_1 + 0.4x_1x_3 + 0.4x_2x_3 $, $m(x) = 1+ x_1x_3 + x_2x_3 + x_2x_4 + h(x_5)$,
\item[] Design IV:  $e(x)= -0.2x_1 + 0.4x_1x_3 + 0.4x_2x_3$, $m(x) = 1+ x_1x_3 + x_2x_3 + x_2x_4 + \widetilde h(x)$,
\end{itemize}
where $h(x_5)= (5\mathbbm 1\left\{x_5=1\right\}-\mathbbm 1\left\{x_5=2\right\}-1)/2$ and  $\widetilde h(x)=((4x_1+1)\mathbbm 1\left\{x_5=1\right\}-\mathbbm 1\left\{x_5=2\right\}-1)/2+\sin(x_4x_5)$, with $\mathbbm1\{\cdot\}$ denoting the indicator function.

We evaluate six inference methods.  Standard \textbf{BART} obtains the posterior of the conditional mean function $m_{\eta}(\cdot)$ using BART, and then averages over the covariates using Bayesian-bootstrap weights. Implementation relies on the R package $\mathsf{BART}$ \citep{sparapani2021nonparametric}: the posterior of $m_{\eta}(\cdot)$ is computed using the function $\mathsf{gbart}$ with argument $\mathsf{type= wbart}$. We draw 2,000 posterior samples after discarding a burn-in of $500$, with the number of trees set to
$T=200$. Default priors from the package are used; see Appendix \ref{appendix:implem} for details.
\textbf{One-step BART} applies the posterior correction of \cite{yiu2023corrected}, which uses posteriors of both $m(\cdot)$
and $\pi(\cdot)$ obtained via BART or its probit variant.\footnote{We implement probit BART by calling $\mathsf{gbart}$ with $\mathsf{type= pbart}$.} 
\textbf{One-step BART$_{\widehat\gamma}$} as given in \eqref{OneStep} is a variant of the one-step BART that uses a pilot estimate for the propensity score $\pi(\cdot)$.\footnote{It can be viewed as assigning a degenerate prior to the propensity score $\pi(\cdot)$.  Its implementation follows Algorithm \ref{algorithm_cf} but without the debiasing term $\widehat{b}_{\eta}^{k}$.}
\textbf {RoBART},  the robust BART method described in Algorithm \ref{algorithm_cf},  combines the BART posterior of $m_{\eta}(\cdot)$, a plug-in estimator for the propensity score $\pi_0(\cdot)$, and the debiasing term $\widehat{b}_{\eta}$ in (\ref{debiased_bay_est}).  RoBART applies cross-fitting with $K=5$ folds. The propensity score is estimated using a probit LASSO that includes all pairwise interactions of $X$'s as regressors. The debiasing term also involves an estimator for $m_0(\cdot)$, which is constructed by the posterior mean of $m_{\eta}(\cdot)$ produced by BART.  We also consider two variants of a frequentist double machine learning estimator \citep{chernozhukov2017double, chernozhukov2018double}:  \textbf {DML--BART}  estimates $\pi(\cdot)$ using a probit LASSO and $m(\cdot)$ using the BART posterior mean;    \textbf {DML--RF} applies the same estimator for $\pi(\cdot)$ but a random forest estimator for $m(\cdot)$. Both DML variants use cross-fitting with $K=5$ folds and report the usual Wald confidence interval based on the sample variance of the estimated orthogonal score. DML--BART is deliberately constructed from exactly the same partition $(I_k)_{k=1}^K$ and the same pilot estimators $(\widehat m^{(-k)},\widehat\pi^{(-k)})$ as RoBART, so that the equivalence of Section \ref{sec:model} applies and the two procedures differ only in how they quantify uncertainty.

Table \ref{tab:simu_p5_probit} reports the Monte Carlo bias, the coverage probability (CP) of the nominal $95\%$ interval estimates, and the average interval length (CIL) for each method.
Standard BART is biased and undercovers in every design, and its coverage remains far below the nominal level when the sample size doubles.  One-step BART reduces the bias by $16\%$ to $28\%$ and raises coverage by $6$ to $20$ percentage points, but the improvement is not sufficient,  especially when the design complexity increases.  The variant One-step BART$_{\widehat\gamma}$ also reduces the bias and improves the coverage at the sample size $n=1000$, but stops doing so at $n=2000$.
RoBART,  on the other hand,  produces substantially smaller bias and close-to-nominal empirical coverage probabilities throughout all designs.  These results align with our theoretical discussion below Theorem \ref{thm:BvM:BART}.  As every design contains additive components depending on more than one covariate,  the sieve of tree ensembles that carries the posterior mass may fail the stochastic equicontinuity condition $\sup_{\eta\in\mathcal{H}^m_n}|\mathbb{G}_n[m_\eta-m_0]|=o_{P_0}(1)$, and the bias term $b_{0,\eta}$ of Theorem \ref{thm:BvM} need not be asymptotically negligible.  By explicitly removing $b_{0,\eta}$,  our RoBART demonstrates robust performance throughout all designs with its advantage over standard and one-step BART increasing with the complexity of nuisance functions.  The cross-fitted nature of RoBART is likely to contribute to its coverage as well through the reduced dependence between the pilot estimators entering the debiasing term and the subsample that generates the posterior draw.
The comparison of the RoBART and DML--BART rows is a direct numerical check of the equivalence established in Section \ref{sec:model}.  Because the two procedures use the same folds and the same pilot estimators, the average of the fold-specific posterior means of RoBART is algebraically identical to the cross-fitted DML. The table confirms this: the two rows report the same bias to three decimals in six of the eight design and sample size cells and differ by $0.001$ in the remaining two, the residual discrepancy being simulation error from approximating the posterior mean by the average of $S$ finite posterior draws.  In terms of uncertainty quantification,  RoBART exhibits better coverage performance than DML--BART in all eight cells, and the gap widens with the complexity of the design, from $0.7$ and $1.4$ percentage points in Designs I and II to $3.0$ and $4.3$ points in Designs III and IV at $n=1000$, and to $3.9$ points in both Designs III and IV at $n=2000$. The price is a moderate increase in the interval length.  The superior coverage performance of RoBART likely reflects its propagation of the BART posterior uncertainty about $m_\eta$ together with the Bayesian-bootstrap variation,  compared with DML's fixed pilot $\widehat m^{(-k)}$.   DML--RF,  which replaces the BART pilot for the conditional mean by a random forest,  produces larger bias in every cell and worse coverage than DML--BART,  except in Design II at $n=1000$ where the two coincide.  Its intervals are longer than those of DML--BART throughout,  and comparable to or longer than those of RoBART except in Design IV at $n=1000$.

\begin{table}
\centering
\caption{Finite-sample performance of BART-based and DML inference methods for the mean response with missing data. Bias, coverage probability (CP) of the nominal $95\%$ interval estimates and average interval length (CIL), computed over 1,000 Monte Carlo replications.}
\vskip.15cm
\label{tab:simu_p5_probit}
{\small
\setlength{\tabcolsep}{2pt}
\renewcommand{\arraystretch}{1.2}
\begin{tabular}{c l ccc| ccc| ccc| ccc}
\toprule
$n$ & Method & \multicolumn{3}{c}{Design I} & \multicolumn{3}{c}{Design II} & \multicolumn{3}{c}{Design III} & \multicolumn{3}{c}{Design IV}\\
\cmidrule(lr){1-2}\cmidrule(lr){3-5}\cmidrule(lr){6-8}\cmidrule(lr){9-11}\cmidrule(lr){12-14}
 & & Bias & CP & CIL & Bias & CP & CIL & Bias & CP & CIL & Bias & CP & CIL\\
\midrule
1000 & BART                    & 0.088 & 0.813 & 0.331 & 0.087 & 0.797 & 0.315 & 0.166 & 0.448 & 0.315 & 0.174 & 0.426 & 0.315 \\
     & One-step BART           & 0.069 & 0.879 & 0.343 & 0.067 & 0.868 & 0.326 & 0.134 & 0.646 & 0.328 & 0.146 & 0.594 & 0.328 \\
       & One-step BART$_{\widehat\gamma}$ & 0.077 & 0.882 & 0.433 & 0.069 & 0.882 & 0.418 & 0.132 & 0.716 & 0.482 & 0.139 & 0.710 & 0.496 \\
     & RoBART                  & 0.010 & 0.948 & 0.421 & 0.007 & 0.947 & 0.407 & 0.029 & 0.920 & 0.471 & 0.041 & 0.918 & 0.480 \\
     & DML--BART               & 0.011 & 0.941 & 0.413 & 0.007 & 0.933 & 0.395 & 0.029 & 0.890 & 0.440 & 0.040 & 0.875 & 0.441 \\
     & DML--RF                 & 0.039 & 0.936 & 0.444 & 0.035 & 0.933 & 0.420 & 0.076 & 0.853 & 0.472 & 0.071 & 0.870 & 0.477 \\
     \midrule
 2000 & BART                   & 0.059 & 0.816 & 0.236 & 0.057 & 0.812 & 0.224 & 0.106 & 0.547 & 0.224 & 0.116 & 0.473 & 0.224 \\
     & One-step BART           & 0.043 & 0.875 & 0.241 & 0.041 & 0.886 & 0.229 & 0.081 & 0.708 & 0.231 & 0.093 & 0.635 & 0.231 \\
      & One-step BART$_{\widehat\gamma}$ & 0.081 & 0.778 & 0.275 & 0.075 & 0.784 & 0.270 & 0.132 & 0.539 & 0.342 & 0.152 & 0.452 & 0.332 \\
     & RoBART                  & 0.014 & 0.932 & 0.266 & 0.011 & 0.941 & 0.261 & 0.020 & 0.941 & 0.334 & 0.031 & 0.927 & 0.320 \\
     & DML--BART               & 0.014 & 0.928 & 0.263 & 0.011 & 0.933 & 0.251 & 0.020 & 0.902 & 0.294 & 0.031 & 0.888 & 0.292 \\
     & DML--RF                 & 0.037 & 0.901 & 0.312 & 0.033 & 0.922 & 0.296 & 0.064 & 0.818 & 0.361 & 0.061 & 0.837 & 0.361 \\
\bottomrule
\end{tabular}}
\end{table}
\subsection{Empirical Application}
We apply the BART-based methods to a subsample of data from the National Health and Nutrition Examination Survey (NHANES) 2007–2008, previously analyzed by \cite{chan2016globally}.\footnote{The data file \texttt{nhanes\_bmi.csv} is publicly available from \cite{ding2023data}, \emph{Replication Data for: A First Course in Causal Inference}, Harvard Dataverse, V4, \href{https://doi.org/10.7910/DVN/ZX3VEV}{https://doi.org/10.7910/DVN/ZX3VEV}.}
The parameter of interest is the average treatment effect (ATE), defined as $\mathbb{E}[Y_i(1)-Y_i(0)]$, of participating in school meal programs ($D$) on body mass index (BMI) for children and youths aged 4--17 years ($Y$). The covariates $X$ include eleven variables: child age, child gender, race dummies (Black and Hispanic), an indicator for family income above $200\%$ of the federal poverty level, indicators for participation in the special supplemental nutrition program and in the food stamp program, an indicator for childhood food security, an insurance coverage dummy, and the age and gender of the survey respondent (an adult in the family). The sample size is $2330$. 

We adapt BART and DML methods from the simulation study to the ATE setting (see Appendix \ref{appendix:other_para} for RoBART for the ATE),  and present the results in Table \ref{tab:bmi}.  As in the simulation study, RoBART applies cross-fitting with $K=5$ folds, and DML--BART is computed from the same partition and the same pilot estimators as RoBART.  In addition to the methods considered in the simulation study,  we replicate two frequentist estimates from \cite{chan2016globally}: IPW refers to the inverse propensity score weighting estimator (H\'ajek-type) and CAL refers to the calibration estimator recommended by \cite{chan2016globally}.\footnote{We take the calibration estimator with exponential tilting as a representative; other versions of calibration estimates reported in \cite{chan2016globally} are similar in terms of the magnitude and confidence interval.}

Table \ref{tab:bmi} shows that all estimated effects of school meal programs on BMI are small relative to the sample mean BMI of $20.11$ (standard deviation $5.42$). In absolute value,  BART,  DML--RF and IPW yield comparatively larger point estimates,  while RoBART,  DML--BART,  CAL and One-step BART lead to estimates closer to zero.  Every $95\%$ interval estimate contains zero, so the sign of the average effect cannot be determined once sampling uncertainty is taken into account.

The application also reproduces, on real data, the numerical equivalence established in Section \ref{sec:model} and documented in the simulations. RoBART and DML--BART report the same point estimate of $0.07$.  The two procedures differ only in how they quantify uncertainty, and here the RoBART credible interval $[-0.36,0.50]$ is marginally shorter than the corresponding Wald interval $[-0.37,0.51]$. Replacing the BART pilot for the outcome regression by a random forest, as in DML--RF,  moves the point estimate to $0.15$.  One-step BART produces the most dispersed posterior and hence the widest interval among the methods considered.  The variant One-step BART$_{\widehat\gamma}$ leads to a narrower credible interval than One-step BART, although it is still wider than those of the other BART-based methods.
 Overall,  our RoBART,  DML--BART and the calibration estimator of \cite{chan2016globally} combine point estimates close to zero with the shortest intervals,  leading to comparatively precise and stable inference.

\begin{table}
\centering
\caption{Bayesian and frequentist ATE estimates of the effect of school meal programs on children's and youths' BMI, $n=2330$.}
\label{tab:bmi}
\vskip.15cm
{\small 
\renewcommand{\arraystretch}{1.2}
\begin{tabular}{ c l c c c}
\toprule
 &  Method & ATE & $95\%$ CI & CIL \\
\midrule
&BART  &  0.15 & $[-0.30, 0.60]$ & 0.90\\
  &One-step BART   & $-0.02$ & $[-0.79, 0.66]$ & 1.45 \\
   &One-step BART$_{\widehat\gamma}$ & 0.04 & $[-0.44, 0.51]$ & 0.95\\
   &  RoBART  &  0.07 & $[-0.36, 0.50]$ & 0.86 \\
  \midrule
  & DML--BART & 0.07 & $[-0.37, 0.51]$ & 0.88\\
  & DML--RF &  0.15 & $[-0.30, 0.60]$ & 0.90\\
  & IPW  & $-0.14$ &  $[-0.62, 0.34]$ & 0.96\\
  & CAL & $-0.04$ & $[-0.48, 0.40]$ & 0.88\\
\bottomrule
\end{tabular}}
\end{table}

\section{Conclusion}
This paper studies semiparametric posterior inference with BART outside the Donsker regime. We show that the natural one-step BART posterior can retain a centering bias and develop RoBART to remove this bias while preserving the posterior variation generated by BART and the Bayesian bootstrap. For the cross-fitted procedure, we establish a Bernstein--von Mises theorem under primitive conditions tailored to BART. Our point estimator, the average of the fold-specific posterior means in the cross-fitted version, coincides exactly with the corresponding DML estimator. This strong equivalence further justifies our correction procedure. Our theoretical and numerical results demonstrate that RoBART provides a theoretically sound posterior distribution for uncertainty quantification around an established robust and efficient frequentist center.


RoBART and DML share the same point-estimation center when constructed using the same folds and nuisance estimators, so their comparison isolates differences in uncertainty quantification. Although the two procedures have the same first-order asymptotic distribution, our simulations show that RoBART achieves coverage closer to the nominal level in finite samples, with the improvement becoming more pronounced as the nuisance functions become more complex. This pattern is consistent with the different ways in which uncertainty is quantified: RoBART propagates posterior uncertainty about the conditional mean function together with Bayesian-bootstrap variation, whereas the conventional DML Wald interval is based on the estimated orthogonal score and treats the estimated nuisance functions as fixed. These differences vanish asymptotically under our conditions but can matter substantially in finite samples.

\section*{Acknowledgments}
Breunig gratefully acknowledges the support of the Deutsche Forschungsgemeinschaft (DFG, German Research Foundation) under Germany's Excellence Strategy -- EXC-2047/1 -- 390685813. Yu gratefully acknowledges the support of JSPS KAKENHI Grant Number 25K05032.

\newpage
\appendix

\section{Proofs of Main Results}\label{appendix:main:proofs}
In the following, we denote the log-likelihood of the conditional probability density function as
	\begin{align*}
		\ell_n(m_\eta)=
		\sum_{i=1}^nR_i\log f_{m_\eta}(Y_i\mid X_i),
	\end{align*}
which depends on the conditional mean function $m_{\eta}(\cdot)$. We use the empirical process and bootstrap process notations by writing $\mathbb{P}_n[h]=n^{-1}\sum_{i=1}^{n}h(Z_i)$ and $\mathbb{P}^*_n[h]=\sum_{i=1}^{n}W_{ni}h(Z_i)$. Recall the definition of the measurable sets $\mathcal H^m_n$ of functions $m_{\eta}$ such that $\Pi(m_{\eta}\in\mathcal{H}^m_n\mid Z^{(n)})\to_{P_0} 1$. We introduce the conditional prior $\Pi_n(\cdot):=\Pi(\cdot \cap \mathcal{H}^m_n)/\Pi(\mathcal{H}^m_n)$. Our BvM theorem makes use of a frequentist estimator $\widehat{\chi}$ as the centering term. Note that this estimator itself is not needed in constructing the Bayesian point estimator or the credible set. All we require is that it admits the linear representation with the efficient influence function. 
For simplicity of notation, we introduce
\begin{align*}
\rho^m(y,x)=y-m(x).
\end{align*}
The frequentist estimator $\widehat\chi$ is required to satisfy the linear representation with the efficient influence function given in \eqref{eif} under $P_0$. 
We set $\widehat{\chi}=\mathbb{P}_n\left[m_0+\widehat{\gamma}\rho^{m_0}\right]$ and indeed due to Lemma C.2 of \cite{BLY2025supplement} this choice is justified since
\begin{align*}
		\widehat{\chi}&=\mathbb{P}_n\left[m_0+\widehat{\gamma}\rho^{m_0}\right]
		=\mathbb{P}_n\left[m_0+\gamma_0\rho^{m_0}\right]+o_{P_0}(n^{-1/2})
		=\chi_0+ \mathbb{P}_n\widetilde \chi_0+o_{P_0}(n^{-1/2}),
\end{align*}
 which is used in the proofs presented below.
	\begin{proof}[Proof of Theorem \ref{thm:BvM}]
As in Theorem 2 in \cite{ray2020causal}, we assume that the propensity score estimator $\widehat{\pi}$  is constructed from observations that are independent of $Z^{(n)}$, as imposed in Assumption \ref{Assump:Rate}. Accordingly, we treat $\widehat\pi$ as a deterministic sequence $\pi_n$ in the sequel and define $\gamma_n(r,x)=r/\pi_n(x)$. Consequently, we can write $\chi_{\eta,  \gamma_n}=\mathbb{P}_n^*[m_{\eta}+\gamma_n\rho^{m_\eta}]$.

 By Theorem 1.13.1 in \cite{van1996empirical}, we can show this by establishing the convergence of the conditional Laplace transform
\begin{align*}
	I_n(t):=	\frac{\iint_{\eta\in\mathcal{H}_n}\exp\left(t\sqrt{n}(\chi_{\eta,  \gamma_n}-\widehat{\chi}-b_{0,\eta}) \right) \mathrm{d}\Pi_W(W^{(n)}|Z^{(n)})e^{\ell_n(m_{\eta})}\mathrm{d}\Pi(m_{\eta})}{\int_{\eta\in\mathcal{H}_n} e^{\ell_n(m_{\eta})} \mathrm{d}\Pi(m_{\eta})}.
\end{align*}
We proceed with the following decomposition:
		\begin{align*}
			\chi_{\eta,  \gamma_n}-\widehat{\chi}=&\mathbb{P}_n^*\left[m_0+\gamma_n\rho^{m_0}\right]-\mathbb{P}_n\left[m_0+\gamma_n\rho^{m_0}\right] +\mathbb{P}_n^*\left[m_{\eta}-m_0-\gamma_n(m_{\eta}-m_0)\right]\\
			=&(\mathbb{P}^*_{n}-\mathbb{P}_n)\left[m_0+\gamma_0\rho^{m_0}\right]
			+(\mathbb{P}^*_{n}-\mathbb{P}_n)\underbrace{\left[m_{\eta}-m_0-\gamma_n(m_{\eta}-m_0)+(\gamma_n-\gamma_0)\rho^{m_0}\right]}_{=:\vartheta_{n,\eta}}\\
			&+\underbrace{\mathbb{P}_n\left[m_{\eta}-m_0-\gamma_0(m_{\eta}-m_0)\right]}_{=b_{0,\eta}}+\underbrace{\mathbb{P}_n\left[(\gamma_0-\gamma_n)(m_{\eta}-m_0)\right]}_{=:r_{n,\eta}},
		\end{align*}
		by the definition of the bias term $b_{0,\eta}$. 
For the remainder term $r_{n,\eta}$, we further express it as
\begin{align}\label{rnTerm}
	r_{n,\eta}&=(\mathbb{P}_n-P_0)\left[(\gamma_0-\gamma_n)(m_{\eta}-m_0)\right]+P_0\left[(\gamma_0-\gamma_n)(m_{\eta}-m_0)\right],\notag \\
	\left|r_{n,\eta}\right|&\leq \left|(\mathbb{P}_n-P_0)\left[(\gamma_0-\gamma_n)(m_{\eta}-m_0)\right]\right|+\frac
	{1}{c}\Vert 1/\pi_n\Vert_{\infty}\Vert \pi_{n}-\pi_0\Vert_{2,\mu_0}\Vert m_{\eta}-m_0\Vert_{ 2,\mu_0},
\end{align}
by employing the Cauchy--Schwarz inequality and the overlap condition in Assumption \ref{Assump:ID}. Hence, we obtain
	$\sup_{\eta\in\mathcal{H}_n}\sqrt{n}|r_{n,\eta}|=o_{P_0}(1)$,
by invoking the stochastic  equicontinuity condition \eqref{ProductSE} imposed in Assumption \ref{Assump:Donsker} for the first term, and the convergence rate conditions in Assumption \ref{Assump:Rate} for the second term on the right-hand side of the inequality (\ref{rnTerm}).

By definition, we can express the first term as $(\mathbb{P}^*_{n}-\mathbb{P}_n)\left[m_0+\gamma_0\rho^{m_0}\right]=(\mathbb{P}^*_{n}-\mathbb{P}_n)\widetilde{\chi}_{0}$. Using the notation of the influence function at $P_0$ given by $\widetilde{\chi}_{0}(z)=m_0(x)+\gamma_0(r,x)\rho^{m_0}(y,x)-\chi_{0}$, we may write
\begin{align*}
			\chi_{\eta,  \gamma_n}-\widehat{\chi}-b_{0,\eta}=& (\mathbb{P}^*_{n}-\mathbb{P}_n)(\widetilde{\chi}_{0} + \vartheta_{n,\eta})+r_{n,\eta}.
		\end{align*}
Thus, the conditional Laplace transform becomes
\begin{align*}
I_n(t)=\frac{\iint_{\eta\in\mathcal{H}_n}e^{\left(t\sqrt{n}(\mathbb{P}^*_{n}-\mathbb{P}_n)(\widetilde{\chi}_{0}+\vartheta_{n,\eta}) \right)} \mathrm{d}\Pi_W(W^{(n)}\mid Z^{(n)})e^{\ell_n(m_{\eta})}\mathrm{d}\Pi(m_{\eta})}{\int_{\eta\in\mathcal{H}_n} e^{\ell_n(m_{\eta})} \mathrm{d}\Pi(m_{\eta})}\times e^{o_{P_0}(1)}.
\end{align*}
Under Assumption \ref{Assump:Donsker} (i), we apply Lemma \ref{lemma:DP} to the function class $\{\widetilde{\chi}_{0}+\vartheta_{n,\eta}:\eta\in\mathcal{H}_n \}$. Given that the estimated propensity score is bounded away from zero in Assumption \ref{Assump:Rate}, the Glivenko--Cantelli property as well as the required moment conditions on the envelope function are satisfied by the corresponding assumptions on the class $\{m_{\eta}:\eta\in\mathcal{H}_n\}$ in Assumption \ref{Assump:Donsker} (i). The conclusion from Lemma \ref{lemma:DP} gives us
		\begin{align*}
	\sup_{\eta\in\mathcal{H}_n}\left|\mathbb{E}\left[e^{t\sqrt{n}(\mathbb{P}_n^*-\mathbb{P}_n)(\widetilde{\chi}_{0}+\vartheta_{n,\eta})}\Bigm|Z^{(n)}\right]-e^{t^2Var_0(\widetilde{\chi}_{0}(Z)+\vartheta_{n,\eta}(Z))/2}\right|	=o_{P_0}(1).
		\end{align*}
Under Assumption \ref{Assump:Rate} we show in Lemma \ref{lemma:SmallVar} that
\begin{equation}
	\sup_{\eta\in\mathcal{H}_n}\left|Var_0\left(\widetilde{\chi}_{0}(Z)+\vartheta_{n,\eta}(Z)\right)-Var_0\left(\widetilde{\chi}_{0}(Z)\right)\right|\to 0.
\end{equation}
We emphasize that the above uniform convergence only requires the influence function $\widetilde{\chi}_{0}+\vartheta_{n,\eta}$ to be in the $P_0$-Glivenko--Cantelli class, not necessarily the $P_0$-Donsker class. Therefore, we have
\begin{align*}
	I_n(t)
	=e^{t^2Var_0(\widetilde{\chi}_{0}(Z))/2}e^{o_{P_0}(1)}\frac{\int_{\eta\in\mathcal{H}_n}e^{\ell_n(m_{\eta})}\,\mathrm{d}\Pi(m_{\eta})}{\int_{\eta\in\mathcal{H}_n} e^{\ell_n(m_{\eta})} \,\mathrm{d}\Pi(m_{\eta})}=e^{t^2Var_0(\widetilde{\chi}_{0}(Z))/2}e^{o_{P_0}(1)},
\end{align*}
which concludes the proof.
\end{proof}	
The original setup in \cite{BLY2022} concerns estimation of the average treatment effect. Our missing-data example can be viewed as applying the proposal of \cite{BLY2022} to the treated (or control) arm only. For concreteness, we adapt the proof of their Corollary 3.1 to show that the feasible estimator $\widehat{b}_{\eta}$ approximates $b_{0,\eta}$ uniformly well in the missing-data problem. 
	\begin{proof}[Proof of Theorem \ref{coro:BvM:debias}]
	It is sufficient to show that
		$\sup_{\eta\in\mathcal{H}_n}\left|b_{0,\eta}-\widehat{b}_{\eta}\right|=o_{P_0}(n^{-1/2})$,
	where 
		$b_{0,\eta}=	\mathbb{P}_n[(\gamma_0-1)(m_0-m_{\eta}) ]$ and $\widehat{b}_\eta=	\mathbb{P}_n[(\widehat \gamma-1)(\widehat m - m_{\eta})]$.
	We make use of the decomposition
	\begin{equation}\label{dec:proof:feasible:drb}
		b_{0,\eta}-\widehat{b}_{\eta}=	\mathbb{P}_n[(\gamma_0-\widehat \gamma)(m_0-m_{\eta})]
		+ \mathbb{P}_n[(\widehat \gamma-1)(m_0- \widehat m) ].
	\end{equation}
	Consider the first summand on the right-hand side of the previous equation.
From Assumption \ref{Assump:Donsker} we infer
\begin{align*}
		\sqrt n\sup_{\eta\in\mathcal{H}_n}&\left|	\mathbb{P}_n[(\gamma_0-\widehat \gamma)(m_0-m_{\eta})]\right|
		\leq 		\sup_{\eta\in\mathcal{H}_n}\left|\mathbb{G}_n[(\gamma_0-\widehat \gamma)(m_0-m_{\eta})]\right|\\
		&\qquad\qquad\qquad\qquad\qquad\qquad+\sqrt n\sup_{\eta\in\mathcal{H}_n}\left|P_0[(\gamma_0-\widehat \gamma)(m_0-m_{\eta})]\right|\\
		&\qquad\leq o_{P_0}(1)+	O_{P_0}(1)\times \sqrt{n}\Vert \pi_0 - \widehat{\pi}\Vert_{2,\mu_0}\sup_{\eta\in\mathcal{H}_n}\Vert m_\eta-m_0\Vert_{2,\mu_0}=o_{P_0}(1),
	\end{align*} 
using Assumption \ref{Assump:Donsker} (ii), the Cauchy--Schwarz inequality and Assumption \ref{Assump:Rate}.
	Consider the second summand on the right-hand side of \eqref{dec:proof:feasible:drb}. 
Another application of the Cauchy--Schwarz inequality and Assumption \ref{Assump:Rate} yields
	\begin{align*}
		 \mathbb{P}_n[(\widehat \gamma-1)(m_0-\widehat m ) ]&= \mathbb{P}_n[(\gamma_0-1)(m_0 -\widehat m )]+o_{P_0}(n^{-1/2})\\
		 &= \mathbb{P}_n\left[\frac{R-\pi_0(X)}{\pi_0(X)}(m_0(X) -\widehat m(X) )\right]+o_{P_0}(n^{-1/2}).
	\end{align*}
Under Assumption \ref{Assump:Rate}, we apply Lemma \ref{lemma:SmallCenter} to get $\mathbb{P}_n\left[\frac{R-\pi_0(X)}{\pi_0(X)}(m_0(X) -\widehat m(X) )\right]=o_{P_0}(n^{-1/2})$. Hence, we obtain  
			$\mathbb{P}_n[(\gamma_0-1)(m_0 -\widehat m)]=o_{P_0}(n^{-1/2})$,
which completes the proof.
\end{proof}


We present a proof of the validity of the $K$-fold posterior cross-fitting procedure. Recall that we partition the sample of size $n$ into $K$ folds of sizes $n_k \asymp n/K$, $k=1,\dots,K$. For each fold $k$, let $\widehat{\chi}^k$ denote the frequentist DML estimator computed on fold $k$, serving as the centering point in the Bernstein--von Mises theorem. Denote by $\ell_{n_k}(m^k_\eta)$ the likelihood function in fold $k$, and let $Z^{(n_k)}$ and $W^{(n_k)}$ denote the observed data and Bayesian-bootstrap weights in fold $k$, respectively.  Let $a\vee b$ denote the maximum of $a$ and $b$.

\begin{proof}[Proof of Theorem \ref{thm:BvM:BART}]
First, we argue that it suffices to check the conditional convergence for each fold-specific posterior in order to get the desired result.
	For each fold $k$, define the posterior Laplace transform
\begin{align*}
	I_{n_k}(t)
	&= \mathbb{E}_\Pi\!\left[\exp\!\left(t\frac{\sqrt{n}}{K}(\check\chi_\eta^k - \widehat{\chi}^k)\right)\,\Big|\,Z^{(n_k)}\right] \\
	&= \frac{\iint \exp\!\left(t\frac{\sqrt{n}}{K}(\chi_{\eta, \widehat \gamma}^k - \widehat{\chi}^k-\widehat{b}_{\eta}^k)\right)\,
		\mathrm{d}\Pi_W(W^{(n_k)}|Z^{(n_k)})\, e^{\ell_{n_k}(m^k_\eta)}\, \mathrm{d}\Pi(m^k_\eta)}
	{\int e^{\ell_{n_k}(m^k_\eta)}\, \mathrm{d}\Pi(m^k_\eta)}.
\end{align*}
Note that $I_{n_k}(t)$ depends on the entire dataset $(Z^{(n_1)},\dots,Z^{(n_K)})$ through pilot estimation, but the sampling from each posterior distribution is made independently across the $K$ folds. Once we verify the high-level assumptions that lead to Theorem~\ref{thm:BvM} and Theorem~\ref{coro:BvM:debias}, for each fixed $t$ in a neighborhood of zero, we get 	
$I_{n_k}(t) \;\to_{P_0}\; \exp\!\left(\textsc v_0 t^2/(2K)\right)$.
Since posterior draws are generated independently across folds, the joint Laplace transform of 
	\begin{equation}\label{eq:conv_general}
				\mathcal{L}_\Pi\!\left(\frac{\sqrt{n}}{ K}(\chi_{\eta, \widehat \gamma}^1 - \widehat{\chi}^1-\widehat{b}_{\eta}^1)\,\big|\,Z^{(n_1)}\right)\ast\ldots\ast\mathcal{L}_\Pi\!\left(\frac{\sqrt{n}}{ K}(\chi_{\eta, \widehat \gamma}^K - \widehat{\chi}^K-\widehat{b}_{\eta}^K)\,\big|\,Z^{(n_K)}\right)
			\end{equation}
factorizes into
\begin{align*}
	\prod_{k=1}^K I_{n_k}(t) 
	\;\to_{P_0}\; \prod_{k=1}^K \exp\!\left(\frac{\textsc v_0 t^2}{2K}\right) = \exp\!\left(\frac{\textsc v_0 t^2}{2}\right).
\end{align*} 		
This limit is precisely the Laplace transform of a centered Gaussian distribution with variance $\textsc v_0$. By Theorem 1.13.1 of \cite{van1996empirical}, this conditional convergence of Laplace transforms implies the distributional convergence in bounded Lipschitz distance. Thus, the posterior law in \eqref{eq:conv_general} converges conditionally to the correct Gaussian limit.

Now, we check the high-level assumptions that lead to our BvM theorem for the fold-specific posterior. In the remaining proof, we suppress the subscript or superscript related to the $k$-th fold to lower the notational burden. The statement is understood to be with respect to any given fold, for $k=1,\ldots,K$.

First, we provide the construction of measurable sets $\mathcal{H}_n$ such that $\Pi(\eta \in \mathcal{H}_n\mid Z^{(n)})\to_{P_0}1$ in Assumption \ref{Assump:Rate}. As we impose the BART-type priors over the conditional mean function, it suffices to find the proper sets $\mathcal{H}_n^m$ for $m_{\eta}$ such that $\Pi(m_\eta \in \mathcal{H}^m_n\mid Z^{(n)})\to_{P_0}1$. Referring to the sieve set defined by equation \eqref{SieveSet}, we work with 
\begin{equation}
	\mathcal{H}_n^m:=\left\{m_{\eta}: m_{\eta}\in \mathcal{M}_n, \Vert m_{\eta}-m_0\Vert_{ 2,\mu_0}\lesssim \varepsilon_n \right\}.
\end{equation}
We prove the following posterior contraction in Lemma \ref{lemma:rate}:
\begin{equation}
	\Pi(m\in\mathcal{M}_n:\Vert m-m_0\Vert_{2,\mu_0}>C_n\, \varepsilon_n\mid Z^{(n)})\to_{P_0}0,
\end{equation}
with $\varepsilon_n=n^{-\alpha/(2\alpha+q_0)}\sqrt{\log n}$ for any $C_n\to\infty$ as $n,p\to \infty$. 
In Lemma \ref{lemma:rate}, we have shown that $\Pi(m_{\eta}\notin \mathcal{M}_n)\to_{P_0} 0$. Combined with Lemma 1 in \cite{ghosal2007nid}, this yields $\Pi(m_{\eta}\notin \mathcal{M}_n\mid Z^{(n)})\to_{P_0}0$. Taken together, we have shown that $\Pi(m_\eta \in \mathcal{H}^m_n\mid Z^{(n)})\to_{P_0}1$.

Under the uniform boundedness assumption of the conditional mean, the requirement for the envelope functions in \eqref{MomentConditions} is satisfied. Regarding the $P_0$-Glivenko--Cantelli property of $\left\{m_\eta-m_0\right\}$ for $m_\eta\in \mathcal{G}_n$, it is sufficient to verify this for the sieve set $\mathcal{M}_n$.
Let $\Vert\cdot\Vert_n$ denote the empirical $L^2(\mathbb{P}_n)$-norm. For $\mathcal{M}_n$ with $\bar{L}_n\sim n\, \varepsilon^2_n/\log n$ and $\bar{s}_n\sim n\, \varepsilon_n^2/\log (p\vee n)$, the proof of Lemma \ref{lemma:rate} shows that $\log N(\varepsilon,\mathcal{M}_n,\Vert\cdot\Vert_{n})\lesssim  n^{q_0/(2\alpha+q_0)}\log n=o(n)$ for any given $\varepsilon>0$. This satisfies the requirement about the growth of the entropy number in Theorem 2.4.3 of \cite{van1996empirical}, which verifies the $P_0$-Glivenko--Cantelli property. 

For the stochastic equicontinuity condition, we apply the multiplier inequality as stated in Lemma \ref{lemma:product}:
\begin{align*}
	&\sup_{\eta\in\mathcal{H}_n}\left|\mathbb{G}_n\left[\left(\gamma_0- \gamma_n\right)(m_\eta-m_0)\right]\right|\\
	&\leq 4\Vert\gamma_n-\gamma_0\Vert_{\infty}\mathbb{E}_0\sup_{\eta\in\mathcal{H}_n}|\mathbb{G}_n(m_\eta-m_0)|+\sqrt{P_0(\gamma_n-\gamma_0)^2}\sup_{\eta\in\mathcal{H}_n}|P_0(m_\eta-m_0)|\\
	&\lesssim \Vert \pi_n-\pi_0\Vert_{\infty}\mathbb{E}_0\sup_{\eta\in\mathcal{H}_n}|\mathbb{G}_n(m_\eta-m_0)|+\Vert \pi_n-\pi_0\Vert_{2, \mu_0}\sup_{\eta\in\mathcal{H}_n}\Vert m_\eta-m_0\Vert_{2, \mu_0}\\
		&\lesssim \Vert \pi_n-\pi_0\Vert_{\infty}\sqrt{n}\varepsilon_n+\Vert \pi_n-\pi_0\Vert_{2, \mu_0}\varepsilon_n=o_{P_0}(1).
\end{align*}
In the last inequality, we have applied the upper bound $
	\mathbb{E}_0\sup_{m_{\eta}\in\mathcal{M}_n(\varepsilon_n)}|\mathbb{G}_n(m_{\eta}-m_0)|\lesssim \sqrt{n}\,\varepsilon_n$, as given in Lemma \ref{lemma:EP}, where $\mathcal{M}_n(\varepsilon_n)$ is defined.
	
	When it comes to the function class with the additive structure, we modify the posterior rate of contraction by $\varepsilon^{add}_n=\sqrt{\sum_{t=1}^{T_0}n^{-2\alpha^t/(2\alpha^t+q^t_0)}\log n}$ for a fixed $T_0<T$. Following the second part of Lemma \ref{lemma:rate}, we have
	\begin{equation*}
		\Pi(m_\eta\in\mathcal{M}_n:\Vert m_\eta-m_0\Vert_{2,\mu_0}>C_n\, \varepsilon^{add}_n\mid Z^{(n)})\to_{P_0}0,
	\end{equation*}
	for any $C_n\to\infty$	as $n,p\to \infty$. The rest of the proof follows similarly to the H\"older continuous case.
\end{proof}

\section{Useful Lemmas and Auxiliary Results}\label{appendix:auxiliary}
\subsection{Useful Lemmas}
The following lemma is in the same spirit as Lemma 9 in \cite{ray2020causal} with one important difference: we do not restrict the range of the function $\varphi$ to $[0,1]$. This is important, as we apply this lemma with $\varphi=\gamma_n-\gamma_0$, which can take on negative values. Accordingly, we use the more general contraction principle from Theorem 4.12 of \cite{ledoux1991banach} instead of Proposition A.1.10 of \cite{van1996empirical}. This allows us to relax the positive range restriction in \cite{ray2020causal}. 
\begin{lemma}\label{lemma:product}
	Consider a set $\mathcal{H}$ of measurable functions $h:\mathcal{Z}\to \mathbb{R}$ and a bounded measurable function $\varphi$. We have
	\begin{equation*}
		\mathbb{E}\sup_{h\in\mathcal{H}}|\mathbb{G}_n(\varphi h)|\leq 4\Vert\varphi\Vert_{\infty}\mathbb{E}\sup_{h\in\mathcal{H}}|\mathbb{G}_n(h)|+\sqrt{P_0\varphi^2}\sup_{h\in\mathcal{H}}|P_0h|.
	\end{equation*}
\end{lemma}

	We now state the following generalization of Theorem 1 from \cite{ray2021dirichlet}, where the function class $\mathcal{G}_n$ containing $g(\cdot)$ can vary with the sample size.
	 We strengthen it following similar moment conditions as in Lemma 11 of \cite{yiu2023corrected}. As discussed by \cite{ray2021dirichlet} on page 2225 therein, this uniformity refers to the marginal posterior distributions of the process $\widetilde{\chi}_{0}+\vartheta_{n,\eta}$ for $\eta\in\mathcal{H}_n$.  It is not about the distributional convergence of this process as a random element in $\ell^{\infty}(\mathcal{H}_n)$.
\begin{lemma}\label{lemma:DP}
	Suppose $\mathcal{G}_n$ is a sequence of separable classes of measurable functions with envelope functions $G_n$, such that
	\begin{equation*}
		\sup_{g\in\mathcal{G}_n}\left|\frac{1}{n}\sum_{i=1}^ng(Z_i)-\mathbb{E}_0[g(Z)]\right|\to_{P_0} 0.
	\end{equation*}
In addition, $\lim_{C\to\infty}\limsup_{n\to\infty}\mathbb{E}_0[G^2_n\mathbb{I}\{G_n^2>C \}]=0$, and $\mathbb{E}_0[G_n^{4}]=o(n)$. Then for every $t$ in a sufficiently small neighborhood of $0$, 
	\begin{equation*}
		\sup_{g\in\mathcal{G}_n}
		\left|\mathbb{E}_0\left[\exp\left(t\sqrt{n}\sum_{i=1}^n (W_{ni}-1/n)g(Z_i)\right)\Bigm\vert Z^{(n)}\right]-\exp\left(t^2Var_0(g(Z))/2\right) \right|\to_{P_0} 0.
	\end{equation*}
\end{lemma}

\subsection{Smaller Order Terms}
The next lemma verifies that the term arising in the debiasing step is asymptotically negligible. Its proof is in the same spirit as bounding the $R_{n,2}$ term in Lemma C.8 of \cite{BLY2025supplement}.
\begin{lemma}\label{lemma:SmallCenter}
Suppose that the pilot estimator computed over some external independent sample converges to the true conditional mean in the $L^2(\mu_0)$-norm, i.e., $\Vert \widehat{m}-m_0\Vert_{2, \mu_0}\to_{P_0} 0$. If, in addition, the true propensity score is uniformly bounded away from zero, then we have
	\begin{equation}
		 \frac{1}{\sqrt n} \sum_{i=1}^n\frac{R_i-\pi_0(X_i)}{\pi_0(X_i)}(m_0 -\widehat m )(X_i)=o_{P_0}(1).
	\end{equation}	
\end{lemma}
	\begin{proof}
		We condition on $(X_1, \ldots, X_n)$, as well as the pilot estimator $\widehat{m}$, which is computed over the external independent sample. We use the fact that  $(R_i-\pi_0(X_i))$ has conditional mean zero. Specifically, this leads to 
		\begin{align*}
			&\mathbb E_0\left[\Big(\frac{1}{\sqrt n}\sum_{i=1}^n\frac{R_i-\pi_0(X_i)}{\pi_0(X_i)}(\widehat{m}(X_i)-m_0(X_i))\Big)^2\,\Big|\, X_1, \ldots, X_n,\widehat{m}\right]\\
			& = \frac{1}{n}\sum_{i=1}^n\big(\widehat{m}(X_i)-m_0(X_i)\big)^2 \frac{Var_0(R_i\mid X_i)}{\pi_0^2(X_i)}
		\end{align*}
		using that $Var_0(R_i\mid X_i)=\pi_0(X_i)(1-\pi_0(X_i))$. Consequently, by the overlap condition $1\lesssim\pi_0(X_i)$ for all $1\leq i\leq n$, we obtain
		\begin{align*}
				\mathbb{E}_0\left[\Big(\frac{1}{\sqrt{n}}\sum_{i=1}^n\frac{R_i-\pi_0(X_i)}{\pi_0(X_i)}(m_0 -\widehat m )(X_i)\Big)^2\right]
			\lesssim  \Vert \widehat{m}-m_0\Vert_{2, \mu_0}^2=o(1),
		\end{align*}
		where the last equality is due to the $L^2(\mu_0)$-convergence for the pilot estimator $\widehat{m}$. Consequently, the result follows from Markov's inequality.
	\end{proof}

For the next result, recall the definition of $\vartheta_{n,\eta}=\left[m_{\eta}-m_0-\gamma_n(m_{\eta}-m_0)+(\gamma_n-\gamma_0)\rho^{m_0}\right]$ introduced in the proof of Theorem \ref{thm:BvM}.
\begin{lemma}\label{lemma:SmallVar}
Let $\pi_n(\cdot)$ be a sequence of functions uniformly bounded away from zero and $\Vert \pi_n-\pi_0\Vert_{2, \mu_0}\to 0$. Uniformly in $\eta\in\mathcal{H}_n$, assume $\Vert m_{\eta}-m_0\Vert_{ 2,\mu_0}\to 0$. 
Then we have
	\begin{equation*}
		\sup_{\eta\in\mathcal{H}_n}|Var_0(\widetilde{\chi}_{0}+\vartheta_{n,\eta})-Var_0(\widetilde{\chi}_{0})|\to 0.
	\end{equation*}
\end{lemma}
\begin{proof}
Since $\widetilde{\chi}_0(Z)$ is centered under $P_0$, we obtain by elementary calculation and the Cauchy--Schwarz inequality
\begin{align*}
\left|Var_0(\widetilde{\chi}_{0}(Z)+\vartheta_{n,\eta}(Z))-Var_0(\widetilde{\chi}_{0}(Z))\right|
	&\leq \left|\mathbb{E}_0(\widetilde{\chi}_{0}(Z)+\vartheta_{n,\eta}(Z))^2-\mathbb{E}_0\widetilde{\chi}_{0}^2(Z)\right|+(\mathbb{E}_0\vartheta_{n,\eta}(Z))^2\\
	&\leq 2\sqrt{\mathbb{E}_0\widetilde{\chi}_{0}^2(Z)}\sqrt{\mathbb{E}_0\vartheta_{n,\eta}^2(Z)}+2\mathbb{E}_0\vartheta_{n,\eta}^2(Z)\\
	&= 2\left(\sqrt{\textsc v_0}\Vert\vartheta_{n,\eta}\Vert_{ 2,P_0}+\Vert\vartheta_{n,\eta}\Vert_{ 2,P_0}^2\right).
\end{align*}
A close inspection of the function $\vartheta_{n,\eta}$ shows that we can proceed by 
\begin{align*}
	\vartheta_{n,\eta}(Z)&=(1-\frac{R}{\pi_n(X)})(m_{\eta}(X)-m_0(X))+(\gamma_n(R,X)-\gamma_0(R,X))(Y-m_0(X))\\
	&=\frac{\pi_n(X)-\pi_0(X)}{\pi_n(X)}(m_{\eta}(X)-m_0(X))+\frac{\pi_0(X)-R}{\pi_n(X)}(m_{\eta}(X)-m_0(X))\\
	&\quad +(\gamma_n(R,X)-\gamma_0(R,X))(Y-m_0(X)).
\end{align*}
We now make use of the assumption that $\pi_n(\cdot)$ is uniformly bounded away from zero, as well as the assumption that the conditional variance of $R_i(Y_i-m_0(X_i))$ is in $L^2(\mu_0)$, i.e., $\int\sigma_0^2(x)\,\mathrm{d}\mu_0(x)<\infty$. Thus, we obtain
\begin{equation*}
	\Vert \vartheta_{n,\eta}\Vert_{ 2,P_0}	\lesssim \Vert m_{\eta}-m_0\Vert_{ 2,\mu_0}+\Vert\pi_n-\pi_0\Vert_{ 2,\mu_0}+\Vert m_{\eta}-m_0\Vert_{ 2,\mu_0}\Vert\pi_n-\pi_0\Vert_{ 2,\mu_0},
\end{equation*}
after some simple algebraic steps. The desired conclusion follows from the convergence of the pilot estimator of the propensity score and $\sup_{\eta\in\mathcal{H}_n}\Vert m_{\eta}-m_0\Vert_{ 2,\mu_0}\to 0$.
\end{proof}

\section{Potential Outcomes and Treatment Effects}\label{appendix:other_para}
As in \cite{ray2020causal} and \cite{BLY2022}, our robust procedure extends naturally to causal inference. Building on the connection between causal inference and missing-data problems \citep{imbens2004}, the additional bias correction terms take the same form as in \cite{BLY2022,BLY2024}.
For individual $i$, consider a treatment indicator $D_i\in\{0,1\}$.
The observed outcome $Y_i$ is determined by 
$Y_i =D_i Y_i(1)  +(1 - D_i)Y_i(0)$ where $(Y_i(1), Y_i(0))$ are the potential outcomes of individual $i$ associated with  $D_i=1$ or $0$. We assume unconfoundedness, i.e., that $(Y_i(1), Y_i(0))$ is independent of $D_i$ given $X_i$, as well as the usual overlap condition.
The covariates for individual $i$ are denoted by $X_i$, a vector of dimension $p$, with the distribution $F_0$ and the density $f_0$. The key functional component is the conditional mean function
$m_0(d,x):=\mathbb{E}_0[Y_i|D_i=d,X_i=x]$.
One of the most commonly used causal parameters is the average treatment effect (ATE):
	$\tau_0:=\mathbb{E}_0[m_0(1,X)-m_0(0,X)]$.
We can also express it in the one-step updated form incorporating the correction term:
\begin{equation*}
	\tau_0=\mathbb{E}_{0}\left[(m_0(1,X)-m_0(0,X))+\gamma^{\mathrm{ATE}}_0(D,X)(Y-m_0(D,X))\right],
\end{equation*}
where its Riesz representer is
\begin{equation*}
	\gamma_0^{\mathrm{ATE}}=\frac{D}{\pi_0(X)}-\frac{1-D}{1-\pi_0(X)}.
\end{equation*}
 
The procedure can be applied separately to the mean potential outcomes under treatment and control and then combined to obtain the ATE, as in \cite{BLY2022} for Gaussian process priors with prior and posterior adjustment. Here, we extend this construction to BART-type priors. The corresponding algorithm for posterior draws of the ATE is given below. 
 Let $\widehat{m}$ and $\widehat{\pi}$ be pilot estimators for the conditional mean and propensity score computed over some external sample. We denote the estimated Riesz representer as
 \begin{equation*}
 	\widehat{\gamma}^{\mathrm{ATE}}=\frac{D}{\widehat\pi(X)}-\frac{1-D}{1-\widehat\pi(X)}.
 \end{equation*}
 Throughout this appendix, let $\boldsymbol{\tau}[\cdot]$ denote the
linear ATE score operator, defined for a candidate function $m(\cdot,\cdot)$
and an observation $z=(y,d,x^\top)^\top$ by
\begin{align}\label{ATEscore}
\boldsymbol{\tau}[m](z)
:=
m(1,x)-m(0,x)
-
\widehat\gamma^{\mathrm{ATE}}(d,x)m(d,x).
\end{align}

\begin{algorithm}[H]
    \caption{Posterior Computation of the ATE}
    \label{algorithm_ATE}

    \For{$s=1,\ldots,S$}{

        (a) Generate the $s$-th posterior draw using the BART prior
        and the data, and denote it by
        $(m_\eta^s(D_i,X_i))_{i=1}^n$.
        \;

        (b) Draw weights $W^s_{ni}=e^s_i/\sum_{j=1}^n e^s_j$ where $e_i^s \stackrel{iid}{\sim} \textup{Exp}(1)$, $1\leq i\leq n$.

        (c) Calculate the recentered posterior draw for the ATE: $\check{\tau}_{\eta}^s:=\tau_\eta^s-\widehat{b}^s_{\eta}$, where
        \begin{align*}
            \tau_\eta^s
            =
            \sum_{i=1}^n W_{ni}^s
            \Big[
                m_\eta^s(1,X_i)-m_\eta^s(0,X_i)
                +
                \widehat{\gamma}^{\mathrm{ATE}}(D_i,X_i)
                \bigl(
                    Y_i-m_\eta^s(D_i,X_i)
                \bigr)
            \Big],
        \end{align*}
        and
        \begin{equation}
            \label{debiased_bay_est_ATE}
            \widehat{b}_\eta^s
            =
            \frac{1}{n}
            \sum_{i=1}^n
            \bm{\tau}
            \bigl[
                m_\eta^s(\cdot,\cdot)
                -
                \widehat{m}(\cdot,\cdot)
            \bigr](Z_i),
        \end{equation}
        with the ATE score operator $\bm{\tau}[\,\cdot\,]$ defined in \eqref{ATEscore}.
                }
    \Output{
        $\left\{
            \check{\tau}_\eta^s
            :
            s=1,\ldots,S
        \right\}$
    }
\end{algorithm}

Besides the ATE, the average treatment effect on the treated (ATT) is of policy interest, especially to economists \citep{heckman1997matching}. In order to facilitate the comparison with \cite{yiu2023corrected}, who also study the ATT, we outline our strategy to model the ATT, which is defined as follows: $\theta_{0}=\mathbb{E}_0[D_i(Y_i-m_0(0,X_i))]/\mathbb{E}_0[D_i]$.
Its one-step updated version is
\begin{equation*}
	\theta_{0}=\frac{\mathbb{E}_{0}[D_i(Y_i-m_0(0,X_i))]}{\mathbb{E}_{0}[D_i]}+\mathbb{E}_{0}\left[\gamma^{\mathrm{ATT}}_0(D,X)(Y-m_0(D,X))\right],
\end{equation*}
where the Riesz representer is given by
\begin{align*}
	\gamma^{\mathrm{ATT}}_0(D,X)=\frac{D}{\bar\pi_0}-\frac{1-D}{\bar\pi_0}\frac{\pi_0(X)}{1-\pi_0(X)},
\end{align*}
where $\bar\pi_0:=\mathbb{E}_{0}[D_i]=\mathbb{E}_{0}[\pi_0(X)]$ denotes the scalar treatment proportion, to be distinguished from the propensity score function $\pi_0(\cdot)$. In this case, the estimated Riesz representer is
\begin{align*}
	\widehat{\gamma}^{\mathrm{ATT}}(D,X)=\frac{D}{\widehat{\bar\pi}}-\frac{1-D}{\widehat{\bar\pi}}\frac{\widehat\pi(X)}{1-\widehat\pi(X)},
\end{align*}
where $\widehat{\bar\pi}=\frac{1}{n}\sum_{i=1}^{n}D_i$. We only need to obtain the pilot estimator and the posterior draws for the conditional mean in the control arm.

Algorithm \ref{algorithm_ATE_cf} describes the $K$-fold cross-fitted version of RoBART for the ATE.  

 \begin{algorithm}[H]
 	\caption{RoBART for the ATE with $K$-fold cross-fitting}
 	\label{algorithm_ATE_cf}
 (a) Take a $K$-fold random partition $(I_{k})_{k=1}^K$ of observation indices. The size of each fold $I_k$ is $n/K$.\;

    \For{$k=1,\ldots,K$}{
  (b) Use the subsample $\{(D_i, X_i^\top)^\top\}_{i\notin I_{k}}$ to construct the estimator $\widehat \pi^{(-k)}(x)$. Set $\widehat{\gamma}^{(-k)}_{\mathrm{ATE}}(d,x)=d/ \widehat \pi^{(-k)}(x)- (1-d)/(1-\widehat \pi^{(-k)}(x))$.\;

 (c) Use the subsample $\{(Y_i, D_i, X_i^\top)\}_{i\notin I_{k}}$ to construct the estimator $\widehat m^{(-k)} (d,x)$.\;

 	    \For{$s=1,\ldots,S$}{
 		(d) Generate the $s$-th draw of the posterior of $(m_\eta(D_i,X_i))_{i\in I_{k}}$
 		 using the BART prior and the subsample $\{(Y_i, D_i, X_i^\top)\}_{i\in I_{k}}$; denote it as $(m^{k,s}_\eta(D_i,X_i))_{i\in I_{k}}$.\;

(e)  Draw Bayesian-bootstrap weights $W_{i}^{k,s} = e_i^{k,s} / \sum_{j\in I_{k}} e_j^{k,s}$ with $e_i^{k,s} \stackrel{\text{iid}}{\sim} \mathrm{Exp}(1)$, $i\in I_{k}$.\;

(f) Calculate the recentered posterior draw for the ATE: $\check{\tau}_{\eta}^{k,s}:=\tau_\eta^{k,s}-\widehat{b}^{k,s}_{\eta}$, where
 		\begin{align*}
 			\tau_\eta^{k,s}= \sum_{i\in I_{k}} W^{k,s}_{i}\left[(m^{k,s}_\eta(1,X_i)-m^{k,s}_\eta(0,X_i))+ \widehat{\gamma}^{(-k)}_{\mathrm{ATE}}(D_i,X_i) \big(Y_i-m^{k,s}_\eta(D_i,X_i)\big)\right],
 		\end{align*}
 		and 
 		\begin{equation}\label{debiased_bay_est_ATE_cf}
 			\widehat{b}^{k,s}_{\eta}=\frac{K}{n}\sum_{i\in I_{k}} \bm{\tau}^{(-k)}[m_\eta^{k,s}(\cdot,\cdot)-\widehat m^{(-k)}(\cdot,\cdot)](Z_i),
 		\end{equation}
 		where $\bm{\tau}^{(-k)}[\,\cdot\,]$ is the ATE score operator \eqref{ATEscore} with $\widehat{\gamma}^{\mathrm{ATE}}$ replaced by the fold-specific $\widehat{\gamma}^{(-k)}_{\mathrm{ATE}}$.
 		
 	}
 	}
 		  \Output{$\{\check{\tau}_\eta^{s}=\frac{1}{K}\sum_{k=1}^K\check{\tau}_\eta^{k,s}:s=1,\ldots,S\}$.} 
 \end{algorithm}

\counterwithin{theorem}{section}

\section{Technical Results for Regression Trees}\label{sec:barttheory}

 In tree-structured models, the idea is to recursively apply binary splitting rules to partition the support of covariates. Although tree-based partitioning allows splits to occur anywhere in the domain, it is often preferable to select split-points from a discrete set \citep{chipman2010bart}. Recall that for the Cartesian product of $p$ subsets of $\mathbb{R}$, i.e., $\Omega\subset\mathbb{R}^p$, we denote the $j$-th projection mapping of $\Omega$ by $[\Omega]_j=\{x_j\in \mathbb{R}:(x_1,\ldots,x_p)^\top\in \Omega \}$. For a given split-net $\mathcal{X}$, we call each point $x_i=(x_{i1},\ldots,x_{ip})^\top$ a split-candidate. For a given splitting coordinate $j$ and a split-net $\mathcal{X}$, a split-point will be chosen from $[\mathcal{X}]_j\cap \textsf{int}([\Omega]_j)$ to dichotomize a box $\Omega$. Note that $[\mathcal{X}]_{j}=\{x_{ij}\in[0,1], i=1, \ldots,b_n \}$ may have fewer elements than $\mathcal{X}$ due to duplication. We denote by $b_j(\mathcal{X})$ the cardinality of $[\mathcal{X}]_j$, i.e., the number of unique elements in the $b_n$-tuple $(x_{1j},\ldots,x_{b_nj})$.

In this appendix, we prove the posterior rate of contraction for BART. The first rigorous analysis covering the random-design case appears in \cite{jeong2023art} for a more general piecewise heterogeneous anisotropic H\"older class. Our setup can be seen as a special case of \cite{jeong2023art} by restricting to the isotropic H\"older continuous class. For two generic probability densities $p$ and $q$, we denote the Kullback--Leibler (KL) divergence by $K(p,q)$ and the square KL variation by $V(p,q)$; see Appendix B in \cite{ghosal2017fundamentals}. Furthermore, denote the Hellinger distance by $\rho_{H}(\cdot,\cdot)$. 
Our derivation of the posterior contraction rate builds on considering sieve sets on which the prior concentrates.

With given $\mathcal{E}$ and $0<\bar{M}<\infty$, we define the function space
\begin{equation*}
	\mathcal{M}_{\mathcal{E},\bar{M}}:=\left\{m_{\mathcal{E},\mathcal{B}}\in \mathcal{M}_{\mathcal{E}}:\Vert \mathcal{B}\Vert_{\infty}\leq \bar{M}\right\},
\end{equation*}
where $\Vert\cdot\Vert_{\infty}$ denotes the supremum norm of the vector $\mathcal{B}$. We consider the collection of all $	\mathcal{M}_{\mathcal{E},\bar{M}}$ such that the number of terminal nodes is upper bounded by $\bar{L}_n$ and the number of active splitting variables is upper bounded by $\bar{s}_n$ for each individual tree in the following sieve set
\begin{equation}\label{SieveSet}
	\mathcal{M}_n:=\mathcal{M}_{\bar{s}_n,\bar{L}_n,\bar{M}}:=\bigcup_{\mathcal{E}:|S|\leq \bar{s}_n,L^t\leq \bar{L}_n,1\leq t\leq T}	\mathcal{M}_{\mathcal{E},\bar{M}},
\end{equation}  
where $S\subseteq\{1,\ldots,p\}$ is the set of splitting variables used by $\mathcal{E}$ (with cardinality $s=|S|$), $\bar{L}_n\sim n\, \varepsilon^2_n/\log n$, $\bar{s}_n\sim n\, \varepsilon_n^2/\log (p\vee n)$, 
and $\bar{M}$ is the upper bound for individual step height in its prior distribution. Because the prior for each step height has a bounded density with compact support, we work with a fixed $\bar{M}$, rather than a growing upper bound for the normal density in \cite{jeong2023art}.

\begin{lemma}\label{lemma:rate}
Under the assumptions stated in Theorem \ref{thm:BvM:BART} for $m_0\in 	\mathcal{F}_p(\alpha,\mathcal{S}_0)$, we have
\begin{equation*}
	\Pi\left(m\in\mathcal{M}_{n}:\Vert m-m_0\Vert_{2,\mu_0}>C_n\, \varepsilon_n\bigm| Z^{(n)}\right)\to_{P_0}0,
\end{equation*}
with $\varepsilon_n=n^{-\alpha/(2\alpha+q_0)}\sqrt{\log n}$ for any $C_n\to\infty$ as $n,p\to \infty$. For $m_0\in \mathcal{F}^{add}_p(\bm{\alpha},\bm{\mathcal{S}}_0)$, the posterior rate of contraction holds with $\varepsilon^{add}_n$.
\end{lemma}
\begin{proof}
We prove the result for $m_0$ in the H\"older continuous class and outline the changes needed for the additive class. Let $m_1,m_2$ denote two generic conditional mean functions in the Gaussian model for the outcome variable with associated densities $f_{m_1}, f_{m_2}$. Then, the Hellinger distance $\rho_H(\cdot,\cdot)$ satisfies
\begin{equation*}
	\Vert m_1-m_2\Vert^2_{2,\mu_0}\lesssim	\rho_{H}^2(f_{m_1},f_{m_2})\lesssim \Vert m_1-m_2\Vert_{2,\mu_0},
\end{equation*}
by Lemma B.1 in the supplementary material of \cite{xie2020adapt} under our Assumption \ref{Assump:BARTPrior} (iii). Hence, it is sufficient to check the convergence in terms of $\rho_{H}$. In addition, we have
\begin{equation*}
	\max\{K(f_{m_0},f_{m_\eta}),V(f_{m_0},f_{m_\eta})\}\lesssim \Vert m_0-m_\eta\Vert^2_{2,\mu_0}.
\end{equation*}
Define $B_n:=\{m_{\eta}: \max\{K(f_{m_0},f_{m_{\eta}}), V(f_{m_0},f_{m_{\eta}})\}\leq \varepsilon_n^2\}$. To check the posterior rate of contraction, we verify the high-level assumptions from Theorem 2.1 in \cite{ghosal2000rates}:
\begin{align}
	&\Pi(m_{\eta}\in B_n)\geq c_1 \exp(-c_2n\, \varepsilon_n^2),\label{RatesBART:eq1}\\
	&\log  N(\varepsilon_n,\mathcal{M}_n,\rho_{H})\leq c_3 n\, \varepsilon_n^2 ,\label{RatesBART:eq2}\\
	&	\Pi(m_{\eta}\notin\mathcal{M}_n)\leq \exp(-c_4n\, \varepsilon_n^2),\label{RatesBART:eq3}
\end{align}
for positive constants $c_1,\ldots,c_4$.

Referring to the first inequality (\ref{RatesBART:eq1}), it follows by the direct calculation of the normal likelihood:
\begin{equation}\label{BnBound1}
	B_n\supset \{m_{\eta}: \Vert m_{\eta}-m_0\Vert_{ 2,\mu_0}\leq C_1\varepsilon_n \},
\end{equation}
under Assumption \ref{Assump:BARTPrior} (iii). Then we construct an approximating ensemble denoted by $\widehat{\mathcal{E}}=\{\widehat{\mathcal{T}}^1,\ldots,\widehat{\mathcal{T}}^T\}$. By restricting the function space to the one constructed by $\widehat{\mathcal{E}}$, we have
\begin{equation}\label{BnBound2}
	\Pi(m_{\eta}:\Vert m_{\eta}-m_0\Vert_{ 2,\mu_0}\leq C_1\varepsilon_n)\geq \underbrace{\Pi(\widehat{\mathcal{E}})}_{\circled{1}} \underbrace{\Pi(m_{\eta}\in \mathcal{M}_{\widehat{\mathcal{E}}}:\Vert m_{\eta}-m_0\Vert_{ 2,\mu_0}\leq C_1\varepsilon_n)}_{\circled{2}}.
\end{equation}
We provide the lower bound for the two prior probabilities separately. 
For a given split-net $\mathcal{X}$, there exists by Assumption \ref{Assump:BARTTree} a tree partition $\widehat{\mathcal{T}}$ generating an approximating function $m_{0,\widehat{\mathcal{T}},\widehat{\bm{\beta}}}$ such that $\Vert m_0-m_{0,\widehat{\mathcal{T}},\widehat{\bm{\beta}}}\Vert_{\infty}\lesssim \varepsilon_n$. An approximating ensemble $\widehat{\mathcal{E}}$ can be constructed by setting $\widehat{\mathcal{T}}^1$ to be $\widehat{\mathcal{T}}$ and $\widehat{\mathcal{T}}^t=[0,1]^p$ for $t=2,\ldots,T$ so that the rest of the trees are root nodes with no splits. Under Assumption \ref{Assump:BARTPrior} (i)(ii), we have
\begin{equation*}
\log \Pi( \widehat{\mathcal{E}})=\sum_{t=1}^T\log \Pi(\widehat{\mathcal{T}}^t)=\log \Pi(\widehat{\mathcal{T}}^1)+(T-1)\log(1-\nu)\geq -C_2n\, \varepsilon_n^2,
\end{equation*}
where the last inequality follows from Lemma 4 in \cite{jeong2023art}.

Referring to the second part $\circled{2}$, we have $\Vert m_\eta-m_0\Vert_{ 2,\mu_0}\lesssim \Vert m_\eta-m_{0,\widehat{\mathcal{T}},\widehat{\bm{\beta}}}\Vert_{\infty}+\varepsilon_n$ for some $m_{0,\widehat{\mathcal{T}},\widehat{\bm{\beta}}}\in \mathcal{M}_{\widehat{\mathcal{T}}}$. We set all trees in $\widehat{\mathcal{E}}$ to be the root nodes except for the first one $\widehat{\mathcal{T}}^1=\widehat{\mathcal{T}}$. Every step-height vector $\mathcal{B}$ for $\widehat{\mathcal{E}}$ has the form $\mathcal{B}=(\bm{\beta}^{1},\bm{\beta}^2,\ldots,\bm{\beta}^T)^\top\in \mathbb{R}^{\widehat{L}+T-1}$ with $\bm{\beta}^1\in \mathbb{R}^{\widehat{L}}$ and $\bm{\beta}^t\in \mathbb{R}$ with $t=2,\ldots,T$, where $\widehat{L}$ is the size of $\widehat{\mathcal{T}}$. By letting $\widehat{\mathcal{B}}=(\widehat{\bm{\beta}},0,\ldots,0)^\top$, we can write $m_{0,\widehat{\mathcal{T}},\widehat{\bm{\beta}}}=m_{0,\widehat{\mathcal{E}},\widehat{\mathcal{B}}}$. Then, we obtain
\begin{equation*}
 \Pi(m_{\eta}\in \mathcal{M}_{\widehat{\mathcal{E}}}:\Vert m_{\eta}-m_0\Vert_{2,\mu_0}\leq C_1\varepsilon_n)\geq  \Pi(m_{\eta}\in \mathcal{M}_{\widehat{\mathcal{E}}}:\Vert m_{\eta}-m_{0,\widehat{\mathcal{E}},\widehat{\mathcal{B}}}\Vert_{\infty}\leq C_2 \varepsilon_n).
\end{equation*}
A similar calculation as in the proof of Lemma 5 in \cite{jeong2023art} leads to the lower bound of the right hand side of the above inequality. By Assumption \ref{Assump:BARTPrior} (iii) where the step-height priors are independent and compactly supported, we can simply replace the Gaussian small-ball probability bounds with the volume of order $C\nu^{\widehat{L}_*}_n$ with $\nu_n=\varepsilon_n \sigma^{-1}_{\max}(A^{-1})/\sqrt{\widehat{L}_*}$ for $\widehat{L}_*=\widehat{L}+T-1$, by lower bounding the volume of $\{\mathcal{B}\in \mathbb{R}^{\widehat{L}_*}:\Vert A\mathcal{B}\Vert_2\leq C\nu_n\}$ for some non-singular matrix $A$. Hence, we have
\begin{align*}
	\log \Pi( \widehat{\mathcal{E}})\geq -C_2n\, \varepsilon_n^2,~~~\log \Pi(m_{\eta}\in \mathcal{M}_{\widehat{\mathcal{E}}}:\Vert m_{\eta}-m_0\Vert_{\infty}\leq C_3\varepsilon_n)\geq -C_4 n\, \varepsilon_n^2.
\end{align*}
The above two inequalities combined with (\ref{BnBound1}) and (\ref{BnBound2}) give us the desired bound in \eqref{RatesBART:eq1}. 

Next, we bound the entropy number in \eqref{RatesBART:eq2}. Under the unit Gaussian error assumption, we can bound the Hellinger distance between $f_{m_1}$ and $f_{m_2}$ by the $L^2(\mu_0)$-distance $\Vert m_1-m_2\Vert_{2,\mu_0}$ by Lemma 2.7 (i) in \cite{ghosal2017fundamentals}. We proceed to upper bound $\log N(\varepsilon,\mathcal{M}_n,\Vert\cdot\Vert_\infty)$ for the stronger supremum norm. Define $\bm{\mathcal{E}}_{S,L^1,\ldots,L^T}$ as the collection of all tree ensembles $\mathcal{E}$ with given $S,L^1,\ldots,L^T$. By construction, we have the upper bound
\begin{equation*}
N(\varepsilon_n,\mathcal{M}_n,\Vert\cdot\Vert_\infty)\leq	\sum_{S:|S|\leq \bar{s}_n}\sum_{(L^1,\ldots,L^T):L^t\leq \bar{L}_n}\sum_{\mathcal{E}\in\bm{\mathcal{E}}_{S,L^1,\ldots,L^T}}N(\varepsilon_n, \mathcal{M}_{\mathcal{E},\bar{M}},\Vert\cdot\Vert_\infty),
\end{equation*}
where we recall that $\bar{L}_n\sim n\, \varepsilon^2_n/\log n$ and  $\bar{s}_n\sim n\, \varepsilon_n^2/\log (p\vee n)$.
For any given $\mathcal{E}$ and $\mathcal{B}_1,\mathcal{B}_2\in \mathbb{R}^{\sum_{t=1}^TL^t}$, we have
\begin{equation*}
	\Vert m_{\mathcal{E},\mathcal{B}_1}- m_{\mathcal{E},\mathcal{B}_2}\Vert_{\infty}=\sup_{x\in[0,1]^p}\left|\sum_{t=1}^T\sum_{l=1}^{L^t}(\beta_{1l}^t-\beta_{2l}^t)\mathbbm1\{x\in\Omega_{l}^t\} \right|\leq \sum_{t=1}^TL^t\Vert \mathcal{B}_1-\mathcal{B}_2\Vert_{\infty}.
\end{equation*}
Following the proof of Lemma 6 of \cite{jeong2023art}, the covering number of the sieve set is thus upper bounded by
\begin{align*}
N(\varepsilon_n,&\mathcal{M}_n,\Vert\cdot\Vert_\infty)\\
\leq &\bar{L}_n^T\sum_{s=1}^{\bar{s}_n}\binom{p}{s}\left(s\max_{1\leq j\leq p}b_j(\mathcal{X})\right)^{T\bar{L}_n}N\left(\varepsilon_n/(T\bar{L}_n),\{\mathcal{B}\in \mathbb{R}^{T\bar{L}_n}:\Vert \mathcal{B}\Vert_{\infty}\leq \bar{M} \},\|\cdot\|_{\infty}\right)\\
\leq &\bar{L}_n^T\bar{s}_np^{\bar{s}_n}\left(\bar{s}_n\max_{1\leq j\leq p}b_j(\mathcal{X})\right)^{T\bar{L}_n}\left(3T\bar{L}_n\bar{M}/\varepsilon_n \right)^{T\bar{L}_n}.
\end{align*}
Under Assumption \ref{Assump:BARTTree} (i), that is,  $\max_{1\leq j\leq p}\log b_j(\mathcal{X})\lesssim\log n$,
the logarithm of the above quantity satisfies
\begin{align*}
\log N(\varepsilon_n,\mathcal{M}_n,\Vert\cdot\Vert_{\infty})&\lesssim
 \log \bar{L}_n+\log \bar{s}_n+\bar{s}_n \log p + \bar{L}_n\left(\log \bar{L}_n+\log \bar{s}_n+\log \log n-\log \varepsilon_n\right)\\
 &\lesssim \bar{s}_n\log p+\bar{L}_n\log n
\end{align*}
using 	$\varepsilon_{n}=n^{-\alpha/(2\alpha+q_0)}\sqrt{\log n}$ and the choices of $\bar{L}_n$ and $\bar{s}_n$ given below the sieve set (\ref{SieveSet}). These choices immediately imply $\log N(\varepsilon_n,\mathcal{M}_n,\Vert\cdot\Vert_{\infty})\lesssim n\, \varepsilon_n^2$.

When it comes to the prior mass outside the sieve set, we follow the proof of condition (2.3) in \cite{rockova2020posterior}. It boils down to checking
\begin{equation*}
	\sum_{t=1}^T\Pi\left(L^t>\bar{L}_n\right)+\Pi\left(S:s>\bar{s}_n\mid L^t\leq \bar{L}_n,t=1,\ldots,T \right)\lesssim e^{-n\, \varepsilon_n^2},
\end{equation*}
with proper choices of $\bar{L}_n$ and $\bar{s}_n$ to be made later. Under Assumption \ref{Assump:BARTPrior} (i)(ii), one can apply Corollary 7 of \cite{rockova2020theory} to get
\begin{align*}
	&\log\Pi\left(L^t>\bar{L}_n\right)\lesssim -\bar{L}_n\log \bar{L}_n\lesssim-\bar{L}_n\log n,
\end{align*}
for $\bar{L}_n=\lfloor Cn\, \varepsilon_n^2/\log n \rfloor$. Under Assumption \ref{Assump:BARTPrior} (i) with $\bar{s}_n=\lfloor Cn\, \varepsilon_n^2/\log(p\vee n) \rfloor$, it follows from the proof on page 50 of \cite{jeong2023art} that
\begin{equation*}
\Pi\left(S:s>\bar{s}_n\mid L^t\leq \bar{L}_n,t=1,\ldots,T \right)\lesssim e^{-n\, \varepsilon_n^2}.
\end{equation*}
This leads to the bound in \eqref{RatesBART:eq3} and completes the proof of the posterior rate of contraction for the H\"older continuous class.

We outline the modifications needed for the additive model. The first difference occurs for the construction of the approximating ensemble $\widehat{\mathcal{E}}$. For each $1\leq t\leq T_0$, there exists a tree partition $\widehat{\mathcal{T}}^t$ generating an approximating function $m_{0,\widehat{\mathcal T}^t,\widehat{\bm{\beta}}^t}$ such that $\Vert m^t_{0}-m_{0,\widehat{\mathcal T}^t,\widehat{\bm{\beta}}^t}\Vert_{\infty}\lesssim \varepsilon^t_{n}$. An approximating ensemble $\widehat{\mathcal{E}}$ can be constructed by setting $\widehat{\mathcal{T}}^t$ as aforementioned with $1\leq t\leq T_0$, and taking the rest of the trees to be root nodes with no splits. The remaining proof about the posterior rate of contraction is similar to the proof of Theorem 7 of \cite{jeong2023art}, and hence omitted.
\end{proof}
The next lemma bounds the local empirical process term without assuming the Donsker property. Compared with the more refined analysis in \cite{han2021set}, the following bound is not sharp. Nonetheless, it is sufficient to show the validity of our debiasing step, under the rate condition $\sqrt{n}\,\varepsilon_{n}\|\widehat\pi_n-\pi_0\|_\infty\to 0$. 
\begin{lemma}\label{lemma:EP}
	Given the sieve set $\mathcal{M}_n$ over the tree ensembles in \eqref{SieveSet} and the posterior rate of contraction $\varepsilon_n$, we have
	\begin{equation}\label{BoundEP}
		\mathbb{E}_0\sup_{m_\eta\in\mathcal{M}_n(\varepsilon_n)}|\mathbb{G}_n(m_{\eta}-m_0)|\lesssim \sqrt{n}\varepsilon_n,
	\end{equation}
	where $\mathcal{M}_n(\varepsilon_n):=\{m_{\eta}\in\mathcal M_n:\Vert m_{\eta}-m_0\Vert_{ 2,\mu_0}\lesssim\varepsilon_n \}$.
\end{lemma}
\begin{proof}
The function class $\mathcal{M}_n(\varepsilon_n)$ recenters $\mathcal{M}_n$ by subtracting $m_0$ and it also  restricts to functions around the truth with a radius of order $\varepsilon_n$. We resort to the following inequality as in Equation (2.6) of \cite{han2021set}:
\begin{equation*}
	\mathbb{E}_0\sup_{m_\eta\in\mathcal{M}_n(\sigma)}|\mathbb{G}_n(m_{\eta})|\lesssim\inf_{0\leq\iota\leq \sigma/2}\left\{\sqrt{n}\iota+\int_{\iota}^{\sigma}\sqrt{\log N_{[]}(\varepsilon,\mathcal{M}_n,L^2(\mu_0))}\,\mathrm{d}\varepsilon\right \}.
\end{equation*}
The entropy integral term $\int_{\iota}^{\sigma}\sqrt{\log N_{[]}(\varepsilon,\mathcal{M}_n,L^2(\mu_0))}\,\mathrm{d}\varepsilon$ comes from $L^2$ chaining with bracketing in the Gaussian regime using Bernstein's inequality, starting from $\sigma$ to $\iota$. The residual term $\sqrt{n}\iota$ comes from the bound $\Vert m\Vert_{1,\mu_0}\leq \Vert  m\Vert_{2,\mu_0}$ towards the end level $\iota$ of the $L^2$ chaining, where $\Vert \cdot\Vert_{1,\mu_0}$ denotes the $L^1(\mu_0)$-norm.

For the bracketing entropy bound, we first bound the entropy number in $L^{\infty}$-norm. If $m_1,\ldots,m_N$ are the centers of $\Vert\cdot\Vert_{\infty}$-balls of radius $\varepsilon$ that cover $\mathcal{M}_n$, one obtains the brackets $[m_j-\varepsilon,m_j+\varepsilon]$. Each bracket has $L^2(\mu_0)$-size of at most $2\varepsilon$; see the proof of Corollary 2.7.2 of \cite{van1996empirical}. Given the entropy number in $L^{\infty}$-norm established in Lemma \ref{lemma:rate}, we also have the upper bound for the bracketing entropy number as $\log N_{[]}(\varepsilon_n,\mathcal{M}_n,L^2(\mu_0))\lesssim n\, \varepsilon_n^2$. By taking $\sigma=\varepsilon_n$ and $\iota=\sigma/2$, we then get the desired upper bound $\mathbb{E}_0\sup_{m_\eta\in\mathcal{M}_n(\varepsilon_n)}|\mathbb{G}_n(m_{\eta})|\lesssim \sqrt{n}\varepsilon_n$.
\end{proof}

\section{Implementation of BART}\label{appendix:implem}
We implement BART using the R package $\mathsf{BART}$ \citep{sparapani2021nonparametric}. Specifically, the posterior of $m_{\eta}(X_i)$ is obtained by applying the function $\mathsf{gbart}$ with the argument $\mathsf{type=wbart}$ for continuous outcomes. We generate 2,000 posterior draws after discarding a burn-in of 500. The number of trees is set to $T=200$. All BART priors follow the default choices recommended in the package. We briefly summarize these priors below and refer readers to Appendix B of \cite{sparapani2021nonparametric} and Section 2 of \cite{chipman2010bart} for further details. Some of the priors specified below differ from those in Assumption \ref{Assump:BARTPrior}, and we make the discrepancies explicit at the end of this appendix.

Consider the model $Y\sim N(m_{\eta}(x), \sigma^2)$ given $X=x$. The conditional mean $m_{\eta}(x)$ is approximated by an additive tree learner $\sum_{t=1}^T\sum_{l=1}^{L^t}\beta_l^t\mathbbm1\{x\in \Omega^t_l\}$, where $\mathcal{T}^t:=(\Omega^t_1,\ldots,\Omega^t_{L^t})$ denotes the structure (partition) of the $t$-th tree of size $L^t$, and $\bm{\beta}^t:=(\beta_1^t,\ldots,\beta_{L^t}^t) \in\mathbb{R}^{L^t}$ denotes the step heights (leaf values) at $L^t$ terminal nodes of the $t$-th tree.  The structure of each binary tree $\mathcal{T}^t$ consists of three components: (i) the number of internal nodes, (ii) the splitting variable at each internal node, and (iii) the splitting value for that variable. The priors are as follows: (i) the probability that a node at depth $\delta$ is internal is $0.95(1+\delta)^{-2}$; (ii) the splitting variable is chosen from $p$ covariates with the proportion vector $\vartheta=(\vartheta_1,\ldots,\vartheta_p)^\top\sim \text{Dir}(\theta/p,\ldots,\theta/p)$, where the hyperparameter $\theta$ is induced from $\theta/(\theta + p)\sim \text{Beta}(0.5,1)$; and (iii) the splitting value is chosen uniformly among the observed values of the splitting variable. Conditional on $\mathcal{T}^t$, each step-height $\beta_{l}^t$ follows a normal prior with mean $\sum_{i=1}^n Y_i/n$ and standard deviation $(\max_i Y_i-\min_i Y_i)/(4\sqrt{T})$. The noise standard deviation $\sigma$ is endowed with inverse $\chi^2$ prior with $\nu=3$ degrees of freedom and scale parameter $\lambda$ chosen so that the $0.9$ quantile of the prior equals $\hat\sigma$, the residual standard deviation from a linear regression of $Y$ on $X$.

The posterior for the additive tree is drawn via a Gibbs sampler. Because the priors on the step-heights $\bm{\beta}^t$ and $\sigma$ are conjugate, conditional on the tree structure $\mathcal{T}^t$, these parameters can be drawn directly: $\beta^{t}_l$ from a normal posterior and $\sigma$ from an inverse $\chi^2$ distribution. Conversely,  conditional on $\bm{\beta}^t$ and $\sigma$, the tree structure $\mathcal{T}^t$ is updated via a Metropolis–Hastings step detailed in Appendix C of \cite{sparapani2021nonparametric}.

To implement the one-step corrected posterior method of \cite{yiu2023corrected} (One-step BART), we need posterior draws of the propensity score function $\pi_{\eta}(x)$. For this, we implement probit BART by calling the function $\mathsf{gbart}$ with $\mathsf{type=pbart}$ in R package $\mathsf{BART}$. The algorithm employs the data-augmentation technique of \cite{albert1993bayesian}, introducing a latent variable $U_i$ with a truncated normal distribution given $(R_i,X_i)$ and the additive tree model $(\mathcal{T}^t, \bm{\beta}^t)_{t=1}^T$:
\begin{eqnarray*}
&& U_i\mid R_i, X_i, (\mathcal{T}^t, \bm{\beta}^t)_{t=1}^T \sim \mathbb{TN}_{A_i}\left(\sum_{t=1}^T\sum_{l=1}^{L^t}\beta_l^t\mathbbm1\{X_i\in \Omega^t_l\}, 1 \right),
\end{eqnarray*}
where $A_i=(0,\infty)$ (positive real line) if $R_i=1$ and $(-\infty,0)$ (negative real line) if $R_i=0$, and $\mathbb{TN}_A$ denotes a normal distribution truncated to $A$. Here the binary variable entering the probit step is the response indicator $R_i$ in the missing data problem, and the treatment indicator $D_i$ in the treatment effect setting of Appendix \ref{appendix:other_para}; we write $U_i$ for the latent variable to avoid a clash with the observation vector $Z_i=(R_iY_i,R_i,X_i^\top)^\top$. 
The latent $U_i$ are then treated as the continuous outcome for BART. Section 4 of \cite{sparapani2021nonparametric} gives more details on BART for binary outcomes.

\spacingset{1.5}

\bibliographystyle{plainnat}
\bibliography{references}
\end{document}